\documentclass[amsmath,amssymb, nobibnotes, aps, prl, showkeys, nofootinbib, floatfix]{revtex4}
\usepackage{graphicx}
\usepackage{dcolumn}
\usepackage{bm}
\usepackage{docs}
\usepackage{amssymb}
\usepackage{amsmath}
\usepackage{amsfonts}       
\usepackage{mathrsfs}
\usepackage{epsfig}
\usepackage{mathtools}
\usepackage{enumerate}
\usepackage{textcomp}
\usepackage{lipsum}
\expandafter\ifx\csname package@font\endcsname\relax\else
 \expandafter\expandafter
 \expandafter\usepackage
 \expandafter\expandafter
 \expandafter{\csname package@font\endcsname}%
\fi

\newcommand{\ltsima} {$\; \buildrel < \over \sim \;$}
\newcommand{\gtsima} {$\; \buildrel > \over \sim \;$}
\newcommand{\lta} {\lower.5ex\hbox{\ltsima}}
\newcommand{\gta} {\lower.5ex\hbox{\gtsima}}

\newcommand{\lsim}{\raisebox{-.4ex}{$\stackrel{<}{\scriptstyle \sim}$}}

\newcommand{\RNum}[1]{\uppercase\expandafter{\romannumeral #1\relax}}

\begin{document}

\title[Multi-transonic spherical flows around MBHs]
{Multi-transonic spherical-type flows around massive black holes: galactic potential induced standing isothermal shocks}

\author{Sananda Raychaudhuri$^{1}$, Shubhrangshu Ghosh$^{2}$ \thanks{Email address: shubhrang.ghosh@gmail.com}, Partha S. Joarder$^{1}$ }
   
\affiliation{$^{1}$ Center for Astroparticle Physics and Space Science, Department of Physics, Bose Institute, \\ 
   Block EN, Sector V, Salt Lake, Kolkata, India 700091.}
\affiliation{ $^{2}$ High Energy $\&$ Cosmic Ray Research Center, University of North Bengal, Post N.B.U, Siliguri, India 734013.}



\begin{abstract}

Accretion on length-scales of interstellar or even intergalactic is particularly relevant, in the context of spherical or quasi-spherical hot mode accretion, supposedly powering low excitation radio galaxies in the maintenance-mode feedback paradigm. In the present study, we aim to analyze such a spherical-type flow around the host active nucleus in the backdrop of a five-component galactic system (SMBH, stellar, dark matter, diffuse hot gas, $\Lambda$), with the principal intent to address the issue of {\it galactic potential induced shock} formation in the flow, that may contain (dissipative) isothermal standing shocks. The present paper is an extension of Raychaudhuri et al. (2018), who conducted a preliminary investigation of such a problem. The galactic potential, not only renders the flow to be multi-transonic in nature, the flow topology resembles `$X\alpha$' and `$\alpha X$' type trajectories of advective flows in the vicinity of the BHs/compact objects. Owing to the influence of the galactic potential, the entire range of galactic mass-to-light ratio ($\Upsilon_B$) allows shock formation in central to the outer radial regions of our wind-type flows, with the strength of those galactic induced shocks found to be comparable to that of the shocks one would expect in the advective flows in the vicinity of a BH. We also observe that the shock parameters remain sensitive to $\Upsilon_B$. We discussed the possible implication of these shocks in the context of radio source dynamics as well as their potential association with flaring in radio jets. Our study also reveals that galactic potential could substantially augment the mass inflow rate. 


\end{abstract}



\maketitle

\section{Introduction}  
\label{1}

Accretion phenomena on length-scales of interstellar or even intergalactic is, perhaps, relevant in quite a number of astrophysical systems (Pittard et al. 2004), particularly pertaining to hot mode accretion powering nearby low luminous, low excitation radio galaxies (LERGs) or those reside at the center of the cool-core clusters, in the maintenance-mode feedback paradigm. Ambient hot X-ray emitting gaseous medium acts as a source for fueling the host active nucleus of these massive elliptical galaxies (e.g., Di Matteo et al. 2003; Best et al. 2006; Allen et al. 2006; Hardcastle et al. 2007; Balmaverde et al. 2008; Narayan \& Fabian 2011; Janssen et al. 2012; Best \& Heckman 2012; Heckman \& Best 2014; Ineson et al. 2015). A giant spherical or quasi-spherical Bondi-type flow would then likely to prevail up to the central supermassive black holes (SMBHs), with length-scales that may exceed well beyond hundreds of parsecs (pc) to kiloparsecs (kpc) (for details see the introduction in Raychaudhuri et al. 2018, hereinafter RGJ18). Thus, in reality, apart from the flow being impacted by central SMBH, the flow would also expected to be influenced by the galactic gravitational 
field. Several evidences have substantiated the idea of this hot mode accretion powering LERGs, prescribing that Bondi-type flow could be 
able to provide sufficient mass supply rate to reasonably power these low-luminous active galactic nuclei (AGNs), and can fairly account for the observed jet power (e.g., Allen et al. 2006; Nemmen et al. 2007; Narayan \& Fabian 2011; RGJ18). A very pertinent question can then naturally arise as to what extent the galactic gravitational potential would then influence the relevant flow dynamics. 

To address this issue, recently, in RGJ18, the present authors conducted a preliminary investigation of such spherical Bondi-type flow\footnote{Although spherical Bondi-type accretion may not be realistic in many situations, however in the context of giant ellipticals, to a fair degree, one can reasonably approximate the flow to be Bondi-type (see introduction in RGJ18).} by taking into account the effect of entire galactic gravitational potential due to different mass components of the host elliptical galaxy [SMBH, stellar, dark matter (DM), diffuse hot gas] in the presence of the repulsive cosmological constant ($\Lambda$) (in the paradigm of $\Lambda$CDM model of cosmology; CDM refers to cold dark matter), in the backdrop of a five-component galaxy model (SMBH + stellar + CDM + diffuse hot gas + $\Lambda$) in a simple framework. To obtain the galactic potential, the authors followed Mamon \& Lokas (2005b; hereinafter ML05b) who modeled the elliptical galaxy as a four mass-component system (see introduction in RGJ18 for more details). The authors investigated the transonic behavior and studied the fluid properties, primarily focusing on adiabatic class of flows. DM distribution has been modeled through a generalized version of (double power-law) NFW (Navarro, Frenk \& White 1995, 1996) type density profile put forwarded by Jing \& Suto (2000), that is being found to render a good fit to simulated DM halos (see RGJ18, and references therein). Our analysis in RGJ18 revealed that galactic potential can substantially alter the spherical flow dynamics, with the flow appearing to be {\it multi-transonic} in nature. We analyzed global flow topology for a few sample cases and explored the possibility of formation of shocks in the flow. For certain cases, quite remarkably, we found standing (Rankine-Hugoniot-type) shock transitions in the central to outer radial locations of the spherical wind-type flows. It has been pointed out in RGJ18 that if in reality, such galactic potential induced shocks can actually emerge in central to outer regions of realistic outflows and jets, could considerably impact the jet terminal speed, and potentially affect the radio source dynamics. Notably, we also found that the galactic potential has a propensity to augment the mass inflow rate. In this context it is intriguing to note that, in most of the earlier studies that have been conducted with the incorporation of additional physical effects in spherical/quasi-spherical accretion, for instance, magnetohydrodynamic (MHD) effects (e.g., Igumenshchev \& Narayan 2002; Igumenshchev 2006) or convective effects (e.g., Igumenshchev \& Abramowicz 2000), or even with the effect of $\Lambda$ on spherical accretion onto isolated BH (e.g., Mach et al. 2013; Ghosh \& Banik 2015), mass supply rate is found to get suppressed (for more details see section 4 of Narayan \& Fabian 2011). 

In essence, our initial study in RGJ18 tentatively suggests that galactic potential can influence the overall dynamics of spherical/quasi-spherical hot mode accretion and consequently the radio-mode (or maintenance-mode) feedback that plausibly operates in these LERGs, however a robust physical understanding of this feedback mode is currently lacking. This then merits for a broader exploration of such effects of galactic potential on spherical/quasi-spherical flow dynamics, particularly emphasizing on the formation of galactic potential induced shocks in the flow. Before venturing on to investigate the influence of galactic potential on more realistic flow dynamics, or considering a more sophisticated galaxy model, it would be tempting to examine whether the findings in our previous study (RGJ18) remain consistent for other classes of flow solutions, and to explore the possible occurrence of other types of shock transitions in the flow. Such a study is expected to throw more light on the dynamics of hot mode accretion, and could be an useful requirement for realistic models of AGN feedback and galaxy evolution. 

In the present paper, we extend the previous work of RGJ18 to {\it isothermal class of flows}. Isothermal accretion onto BHs have been investigated on numerous occasions in the past, primarily with the motivation of addressing the issue of shock formation in BH accretion and wind (e.g., Chakrabarti 1989a, 1990a; Yang \& Kafatos 1995; Foglizzo 2002; Das et al. 2003). Issue of multi-transonicity and shocks in the context of BH accretion and wind have been extensively examined in the astrophysical literature by several authors, particularly by Chakrabarti and collaborators (e.g., Chakrabarti 1989a,b,c,d; Abramowicz \& Chakrabarti 1990; Chakrabarti \& Wiita 1992; Chakrabarti \& Molteni 1993; Molteni et al. 1994; Chakrabarti 1996; Das 2002). In the context of self-gravitating thick accretion disks, Chakrabarti (1989d) showed the possibilities of having multiple shocks in the rotating wind solutions. Chang \& Ostriker (1985) have obtained solutions for standing shocks in spherical  accretion. Theuns \& David (1992) have investigated spherically symmetric, polytropic flows around a radiating star and showed the possibilities of having shocks in the flow. Moreover, in the context of solar and stellar winds, shock solutions have been obtained previously by many authors (e.g., Axford \& Newman 1967; Axford \& Ip 1986; see Chakrabarti 1989a and references therein). Lee at al. (1977) 
analyzed discontinuous transitions in current-carrying two-fluid plasma. Ferrari et al. (1985), on the other hand, while studying transonic wind-type flows in the context of astrophysical jets obtained possible shock solutions in isothermal wind-type flows. Foglizzo (2002) have studied the linear stability of shocked accretion in isothermal Bondi flow. The characteristic features of such shocks in isothermal flows are markedly different from that of the dissipationless (Rankine-Hugoniot) shocks that we focused in our previous study. Unlike Rankine-Hugoniot shocks which are radiatively very inefficient where the energy of the flow across the shock remains uniform, in the isothermal shock transition, on the other hand, energy of the flow is very efficiently radiated away at the shock location by keeping the temperature distribution of the flow uniform across the shock, maintaining the regular flow structure through the shock (e.g., Chakrabarti 1989a; Chakrabarti 1990b; Abramowicz \& Chakrabarti 1990; Das et al. 2003; Das et al. 2009). Studies of isothermal shocks had been undertaken by various authors, both in the context of isothermal and also adiabatic class of flows (e.g., Abramowicz \& Chakrabarti 1990; Lu \& Yuan 1997, 1998; Fukumura \& Tsuruta 2004), as well as in the context of advection dominated accretion flows (ADAFs) (e.g., Das et al. 2009). It has been pointed out by Das et al. (2009) that isothermal shocks provide the most effective mechanism to release energy from the flow, and may be linked to observed X-ray flares or QPOs in Galactic binaries (Das 2003; Das et al. 2003), or iron fluorescence lines observed in some AGNs (Fukumura \& Tsuruta 2004). Moreover, it has been reported earlier that unlike other shock cases for which the flow thickness changes abruptly at the shock and it is impossible to furnish the accurate pressure balance condition at the shock position, it is only when the flow remains isothermal at the shock, the flow thickness remains the same and one can obtain a correct pressure balance condition (e.g., Chakrabarti 1989a; Das et al. 2003). 

Motivated by these earlier works, here we aim to perform a comprehensive study of spherical Bondi-type accretion for isothermal flows in the backdrop of a five-component elliptical galaxy model adopted in RGJ18, with the principal intent to explore the issue of {\it galactic potential induced (dissipative) isothermal shocks} in spherical flows. We also investigate in detail, how the gravitational potential of the host elliptical would influence the overall dynamics of the flow in comparison to the classical Bondi solution, to check, to what extent our findings in the context of adiabatic flows (in RGJ18) remain consistent for isothermal class of flows. It needs to be mentioned that in RGJ18, we provided a brief introduction to spherical isothermal flows in the context of our galactic system, however, without venturing into any substantive analysis. The original purpose of the present study is to lay the basis for more realistic modeling of accretion and outflow/jet dynamics, and subsequent feedback energetics in the context of these massive galaxies in the present-day Universe. In the following section we briefly describe the hydrodynamical model for isothermal spherical accretion, in our context.

\section{Hydrodynamical spherical accretion model for isothermal flows}
\label{2}

For an isothermal flow, the equation of state follows a simple linearized relation $P = c_s^2 \rho$, where $P$, $\rho$ and $c_s$ are the pressure, density and sound speed of the accreting 
fluid, respectively. $c_s$ is related to the temperature $T$ of the flow through the expression $c_s^2 =  {k_B T}/{\mu m_p}$, where, $k_B$, $m_p$, and $\mu = 0.592$ are the usual Boltzmann constant, proton mass, and the mean molecular weight for the galactic abundance of Hydrogen and Helium; Hydrogen mass fraction being $X = 0.75$, respectively. Since we are primarily interested in quasi-stationary, spherical accretion towards the central SMBH, we express all flow variables only as functions of $r$. Also, as usual, we assume the flow to be inviscid in nature. Throughout in our study, the radial coordinate $r$ is scaled in units of $r_g = {GM_{\rm BH}}/c^2$; $M_{\rm BH}$ is the mass of the central SMBH, $G$ the universal gravitational constant, and $c$ is the speed of 
light. Also, the radial flow velocity $u$ and the isothermal sound speed $c_s$ are scaled in units of $c$. The basic conservation equations for the inviscid, steady state, spherical accretion are well established in the literature. Here, we simply furnish them for the sake of completeness. \\

(i) Mass transfer: 
\begin{eqnarray}
\dot M =  - 4 \pi r^2 \rho \, u \, ,  
\label{1}
\end{eqnarray}
where, $\vert \dot M \vert$ is the Baryon mass accretion rate. \\

(ii) Radial momentum transfer: \\
\begin{eqnarray}
u \frac{du}{dr} + \frac{1}{\rho} \frac{dP}{dr} 
+ \frac{d\Psi_{\rm Gal} \, (r)}{dr} = 0 \, ,
\label{2}
\end{eqnarray}
where, $\Psi_{\rm Gal} \, (r)$ is the net galactic potential; subscript `Gal' signifies `galactic'. Following earlier works to study 
transonic accretion around BH/compact objects 
(e.g., Chakrabarti 1989a; Chakrabarti 1990a,b; Chakrabarti 1996; Narayan et al. 1997; Mukhopadhyay \& Ghosh 2003), combining Eqns. (1) and (2), we obtain 
\begin{eqnarray}
\frac{du}{dr} = \frac{{2 c^2_s}/{r} - \mathscr{F}_{\rm Gal} (r)}{u^2 -c^2_s/u} = \frac{N \, (u, c_s, r)}{D \, (u, c_s)}  \, ,
\label{3}
\end{eqnarray}
where, $\Psi_{\rm Gal} \, (r) = \int \mathscr{F}_{\rm Gal} (r) \, dr$; $\mathscr{F}_{\rm Gal} (r)$ is the net gravitational force function of the host galaxy. 
Here `$N \, (u, c_s, r)$' and `$D \, (u, c_s)$' represent the numerator and 
denominator, respectively, of Eqn. (3). Note that, for an isothermal flow, the flow dynamics can be described solely by Eqn. (3). To ensure that 
the flow remains continuous everywhere, if the denominator of 
Eqn. (3) vanishes at any radial location, the numerator must also vanish at the location. Thus at that particular radius named as `critical radius', or in our case the `sonic radius' ($r_c$), 
the following identity $N=D=0$ should be satisfied; at $r_c$, one then obtains

\begin{eqnarray}
u_{c} = c_{\rm sc} =  \sqrt{\frac{ r_c \, \mathscr{F}_{\rm Gal} \vert_{r_c}}{2}}\, ,
\label{4}
\end{eqnarray}
where $u_{c}$ and $c_{\rm sc}$ are the corresponding radial velocity and sound speed at $r_c$, respectively. To obtain the solution of Eqn. (3), 
one has to apply l'Hospital's rule to Eqn. (3) at sonic location $r_c$, given by  
\begin{eqnarray}
\left.\frac{du}{dr} \right\vert_{r_c} = \pm \frac{1}{\sqrt{2}} \, \left(- \frac{2 c^2_{sc}}{r^2_c} -  \left.\frac{d \mathscr{F}_{\rm Gal} (r)}{dr} \right\vert_{r_c}  \right)^{1/2} \, , 
\label{5}
\end{eqnarray}
where, the ‘$-$’ sign in Eqn. (5) represents the accretion solution, whereas the ‘+’ sign represents the wind solution. 

For an isothermal flow for which the temperature $T$ remains constant throughout the flow regime, given a value of temperature $T$, or equivalently, if the temperature at the outer accretion boundary radius $r_{\rm out}$ is known, one can determine $r_c$ using the relation $T \equiv T_c \equiv T_{\rm out} =  {\left(\mu m_p \, r_c \, \mathscr{F}_{\rm Gal} \vert_{r_c} \right)}/{2 k_B} $, and consequently $u_{c}$ using Eqn. (4). Here, $T_c$ represents the temperature at the sonic location, $T_{\rm out}$ is the temperature at $r_{\rm out}$ or the ambient temperature. Using the fourth order Runge-kutta method, we then solve Eqn. (3) integrating from $r_c$ inwards and outwards. For spherical isothermal flow, the whole problem can thus be resolved, if the temperature $T$ of the flow is solely known (e.g., Das et al. 2003). 

To proceed further, one needs to know the elliptical galaxy potential $\Psi_{\rm Gal} \, (r)$, or the corresponding force function $\mathscr{F}_{\rm Gal} \, (r)$. Following ML05b, RGJ18 made a detailed 
formulation of such a galactic potential for a five-component system (see section 2 in RGJ18). For the sake of completeness, we briefly comment on our elliptical galaxy model in `Appendix A', where we mainly focus on relevant galactic and cosmological parameters that are explicitly required for our computation. Readers are advised to see `Appendix A', before following the rest of the paper. 


\section{Transonic behavior of spherical flows} 
\label{3}

The primary attribute of the classical Bondi solution, described for a spherically symmetric, polytropic\footnote{Here, when we refer polytropic flows, we mean both adiabatic and isothermal class of flows.} accretion flow onto a single point gravitating source, is that, the flow always exhibits uni-transonic behavior. Nonetheless, there are scenarios being reported in the literature that even for spherical accretion onto a central point source, more than one sonic (critical) point can appear in the 
flow. For instance, Chang \& Ostriker (1985) have shown that if some local heating and cooling effects are incorporated, one may obtain more than one critical point in the spherical accretion, where the authors even obtained solutions for standing shocks in the flow. Similarly, Nobili et al. (1991), while studying spherical accretion onto an isolated BH with the inclusion of detailed radiative processes, found a second critical point in the flow. In fact, Turolla \& Nobili (1988), and Nobili \& Turolla (1988), have discussed the possibilities of having multiple critical points in spherical accretion. 

\begin{figure*}
\centering
\includegraphics[width=170mm]{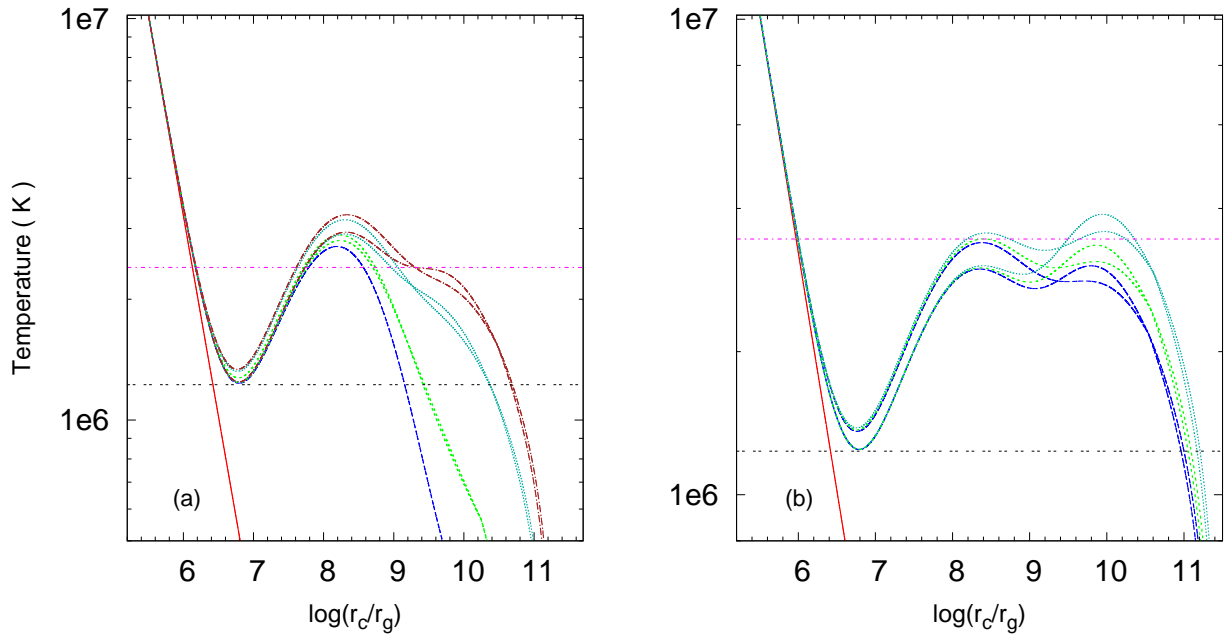}
\caption{Figure 1 shows the variation of temperature $T$ as a function of sonic location $r_c$, for different values of galactic mass-to-light ratio ($\Upsilon_B$). Solid ({\it red online}) curves in both the panels correspond to classical Bondi solution onto an isolated BH. In panel (a), thick curves [short-dashed ({\it green online}), dotted ({\it cyan}), and long dotted-dashed ({\it brown})] correspond to $\Upsilon_B \equiv (33, 100, 150)$, respectively, for JS-3/2 DM profile, while the corresponding thin curves are for NFW DM profile. Long-dashed curve ({\it blue online}) correspond to $\Upsilon_B  = 14$ with no DM 
case. In Fig. 1a, the upper horizontal short dotted-dashed ({\it pink online}) line is for constant temperature $T \simeq 2.4 \times 10^{6} \, K$, intersecting the respective curves at three sonic locations. Panel (b) is similar to that of panel (a), however depicted corresponding to those values of $\Upsilon_B$, for which one can have both three sonic points as well as five sonic points in the flow. In panel (b) thick curves [long-dashed ({\it blue online}), short-dashed 
({\it green online}), and dotted ({\it cyan online})] correspond to $\Upsilon_B \equiv (250, 300, 390)$, respectively, for JS-3/2 DM profile, while the corresponding thin 
curves are for NFW DM profile. Here again, upper horizontal short dotted-dashed ({\it pink online}) line is for constant $T \simeq 3.45 \times 10^{6} \, K$, showing that 
it intersects the curve corresponding to $\Upsilon_B = 390$ with JS-3/2 DM case, at five sonic locations. The lower horizontal double-dashed lines in figures 1a,b 
are representative lines for temperature $T_{\rm min}$. 
 } 
\label{Fig1}
\end{figure*}

Our goal here is to examine how the galactic potential affect the transonic behavior of {\it isothermal} spherical flow. Transonic behavior of an isothermal flow can be described by $T - r_c$ profile, as depicted in Fig. 1. In RGJ18, we discussed the possibility of the emergence of multi-transonicity in spherical isothermal flows in the presence of galactic 
potential, as is being revealed in Fig. 1. The nature of the profiles reflect that depending on the value of galactic mass-to-light ratio $\Upsilon_B$, for a certain range of $T$, say from some $T_{\rm min}$ to $T_{\rm max}$, not only multi-transonicity emerges in the fluid flow for the entire range of $\Upsilon_{B}$, for relatively higher values of $\Upsilon_B$ ($\Upsilon_B > 200$ for NFW DM case; $\Upsilon_B > 240$ for JS-3/2 DM case), even five sonic points appear in the flow for a certain range in the temperature (see Fig. 1b). For clarity, we have depicted representative lines of constant temperatures (short dotted-dashed lines) corresponding to three sonic point case in Fig. 1a, and corresponding to five sonic case in Fig. 1b, respectively. Note that, the sonic points associated with negative slopes of the curves are referred to as the `saddle-type' (or `X-type') sonic points, while in between X-type sonic points, one associated with positive slope of the curves is referred to as `center-type' (or `O-type') sonic point. In figures 1a,b, we also marked $T_{\rm min}$; as an example, in Fig. 1a we show it corresponding to cases $\Upsilon_B = 150$ with NFW profile, and $\Upsilon_B = 14$, represented by the lower horizontal double-dashed line of constant $T \equiv T_{\rm min} \simeq 1.225 \times 10^{6} \, K$, and in Fig. 1b we show it corresponding to cases $\Upsilon_B \equiv (250, 300, 390)$ with NFW DM profile, again represented by lower horizontal double-dashed line of constant $T \equiv T_{\rm min} \simeq 1.235 \times 10^{6} \, K$. In general, corresponding values of $T_{\rm min}$ are different for different choices of $\Upsilon_B$ and corresponding DM profiles. For $T$ less than the corresponding values of $T_{\rm min}$, the flow is devoid of any inner sonic point. Similarly, note that corresponding values of $T_{\rm max}$ are also different for different values of $\Upsilon_B$ and DM  profiles. Figure 1 also reveals that with the enhancement in the value of $\Upsilon_B$ outer X-type sonic location shift outwards, while the inner X-type sonic point almost remains unaltered (see Fig. 1).  

Here it needs to be pointed out that for accretion onto a single point compact gravitating source, a similar nature of multi-transonicity could appear in an (advective flow) close to the BH/compact object, if the flow has sufficiently large angular momentum ($\lambda$) which can provide the requisite centrifugal barrier close to the central object (e.g., Fukue 1987; Abramowicz \& Chakrabarti 1990; Chakrabarti 1990b; Chakrabarti 1996; Mukhopadhyay \& Ghosh 2003). For spherical polytropic flows, on the other hand, instead of $\lambda$, the galactic potential provides the required centrifugal barrier to the flow in the central to outer radial regions where it dominates (see RGJ18 for more details). The emergence of multi-transonicity in the fluid flow signifies the possibility of occurrence of shocks in the flow. Here, it is worth mentioning that, while in the usual advective isothermal flows the global parameter space is described by $\left <\lambda, T \right > $ (Das et al. 2003), in the presence of galactic potential, our global parameter space for isothermal flows is described by $\left < \Upsilon_{B}, T \right >$. In one of the later figures while dealing with the issue of shocks, we show the parameter space region for multi-transonic accretion and wind. 

In the next section, we investigate in details the flow topology and examine the issue of shock formation in the context of our spherical flow. Before proceeding to the next section, we briefly furnish the expressions relating to shocks and allied quantities.    


\subsection{Basic shock quantities}
\label{3.1}

In the context of accretion related phenomena, most of the studies related to the issue of shocks have been predominantly focused on non-dissipative Rankine-Hugoniot type shocks, in which case the energy of the flow across the shock remains uniform, while the temperature of the flow changes perpetually at the shock. Nonetheless, here we would like to direct our attention on another class of shocks, in which case the temperature distribution remains uniform across the shock but the flow is allowed to dissipate energy at the shock; importance of such isothermal shocks has been discussed previously in section I. The necessary conservation conditions at the shock location ($r_{\rm sck}$) are then given by (e.g., Chakrabarti 1989a,b; Abramowicz \& Chakrabarti 1990; Das et al. 2003)

\begin{eqnarray}
\rho_+ u_+ r^2_{\rm sck} = \rho_- u_- r^2_{\rm sck} \, , 
\label{6}
\end{eqnarray}
and 
\begin{eqnarray}
p_+ + \rho_+ u_+^2 = p_- + \rho_- u_-^2 \, , 
\label{7}
\end{eqnarray}
where subscripts `$-$' and `$+$' defines the pre-shock and post-shock quantities. 
Using Eqns. (6-7) and equation of state at isothermal shock location, one can finally arrive at the shock invariant quantity, given by (see Chakrabarti 1989a,b)
\begin{eqnarray}
M_+ + \frac{1}{M_+} = M_- + \frac{1}{M_-} \, , 
\label{8}
\end{eqnarray}
where $M = u/c_s$ is the Mach number of the flow. In order to have shocks in the flow, it is required to satisfy the relations from Eqns. (6-8) simultaneously. 

Isothermal shocks can occur in an accretion flow if the accreting matter has a possibility to jump from outer to the inner sonic point branch, provided 
that the accretion solution traversing through the inner X-type sonic point (denoted by say $r_{I}$) has lower energy content as compared to that for the solution through 
outer X-type sonic point (denoted by say $r_{O}$); here subscripts `$I$' and `$O$' are for inner and outer sonic points, respectively (see Fig. 1a). Conversely, for wind shocks, matter has to jump from the inner to the outer sonic point branch, with  
the energy content of the wind solution at $r_{O}$ should be lower than that at $r_{I}$ (e.g., Abramowicz \& Chakrabarti 1990; Das et al. 2003). Specific energy of the flow can be obtained by integrating Eqn. (2), given by 

\begin{eqnarray}
\mathscr{E} = \frac{u^2}{2} + c^2_s \, \ln(\rho) + \Psi_{\rm Gal} \, .
\label{9}
\end{eqnarray}
The energy difference between $r_{I} $ and $r_{O}$ is then given by 
\begin{eqnarray}
\delta \mathscr{E} =  \mathscr{E}(r_{O}) - \mathscr{E}(r_{I}) =  \, - c_s^2 \, \ln\left(\frac{r_{\rm out}}{r_{\rm in}}\right)^2  + \, \Psi_{\rm Gal} \, (r_{\rm O}) - \Psi_{\rm Gal} \, (r_{I})  \, .
\label{10}
\end{eqnarray}
For accretion shocks $\delta \mathscr{E}$ should be positive, whereas for wind shocks $\delta \mathscr{E}$ should be negative. As we noted, that for isothermal shocks, energy is required 
to be dissipated at the shock location. In terms of pre-shock quantities, the energy dissipation at the shock then amounts to (e.g., Chakrabarti 1989a)
\begin{eqnarray}
\Delta \mathscr{E}= \frac{c_s^2}{2} \left[ M_-^2 - \frac{1}{M_-^2} - 2 \ln(M_-)^2 \right] \, , 
\label{11}
\end{eqnarray}
where, in the context of isothermal shocks, $\Delta \mathscr{E}$ should be positive at the shock location.

\section{Flow topology} 
\label{4}

$T - r_c$ profiles in Fig. 1 indicate that flow would likely to have a complex topology. The flow topology can be described through the radial distribution of the 
Mach number $M$. We extensively analyze the global parameter space spanned by $\left <\Upsilon_B, T \right>$, which reveals that four distinct types of flow topologies are possible pertaining to our spherical flow. For $T$ greater than some $T_{\rm max}$ (see Fig. 1, and second paragraph of the previous section), the flow exhibits 
uni-transonic behavior, where the inner X-type sonic point only provides the possible physical path for the flow to make a sonic transition, resembling the scenario of classical Bondi case. To exemplify, we show the corresponding flow topology in Fig. 2 for various values of $\Upsilon_B$ corresponding to a typical choice of $T \sim 6.5 \times 10^6 \, K$, for few sample cases.

Figure 3 depicts the percentage deviation of accretion and wind velocities associated with our five-component case from that of the classical Bondi case as a function of radius $r$, represented by quantity $\xi = \frac{\mathscr{Q}_{\rm Gal} - \mathscr{Q}_{\rm BH}}{\mathscr{Q}_{\rm BH}} \times 100$; $\mathscr{Q}_{\rm Gal}$ corresponds to any relevant physical quantity for the five-component 
galaxy, while, $\mathscr{Q}_{\rm BH}$ denotes the identical physical quantity for the classical Bondi case. The figure reveals that galactic potential could substantially enhance the accretion flow velocity in the central to outer radial regions, with the accretion velocity can increase by several 100 percent at $r \gtrsim 10^8 \, r_g$. 
This seems to indicate that, to have a spherical/quasi-spherical accretion flow onto central SMBHs originating from beyond the Bondi radius in the presence of our galactic potential, outer boundary value for radial velocity may need to be much higher, as compared to the classical Bondi case. Nonetheless, wind velocity do not show much deviation from that of classical Bondi case. Moreover, contrary to the case for accretion velocity, galactic potential reduces the wind velocity except for the region where $\Lambda$ dominates. Figure 3 also shows that for higher values of $\Upsilon_B$, galactic potential has a greater effect on accretion and wind solutions. We also found from our analysis that the quantity $\xi$ decreases, with the increase in $T$.

\begin{figure}
\centering
\includegraphics[width=90mm]{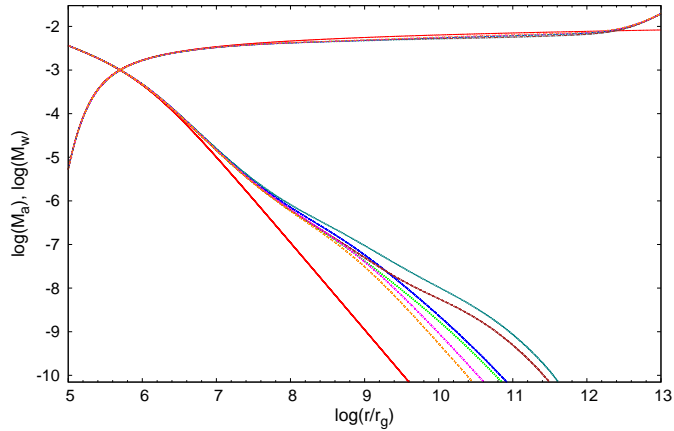}
\caption{Radial profile of Mach number ($M$) for uni-transonic regime where accretion/wind flows traverse only through the inner `X-type' sonic point. Thick curves (below in the figure) are for accretion branch, while thin curves (top in the figure) are for wind branch. Solid curve correspond to, classical Bondi case ({\it red online}), while the long-dashed ({\it blue online}), dotted ({\it cyan online}) and short dotted-dashed ({\it pink online}) curves are for $\Upsilon_B \equiv (100, 390, 33)$ with JS-3/2 DM profile, respectively, and short dashed ({\it green online}), long dotted-dashed ({\it brown online}) are for $\Upsilon_B \equiv (100, 390)$ with NFW DM profiles, respectively. 
Double dashed ({\it golden online}) lines are to represent the case for  $\Upsilon_B = 14 $ for no dark matter case. The corresponding typical value of temperature we choose to generate the profiles is $T \sim 6.5 \times 10^6 \, K$.
 }
\label{Fig2}
\end{figure}

\begin{figure*}
\centering
\includegraphics[width=160mm]{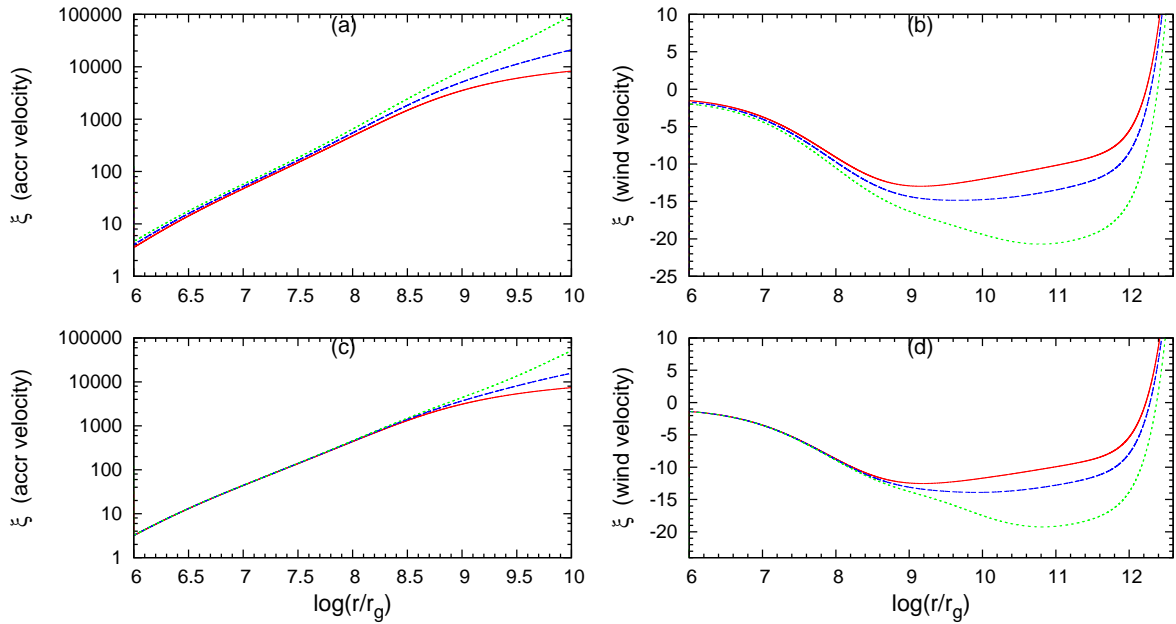}
\caption{Variation of $\xi$ as a function of $r$ for few specific values of $\Upsilon_B$. Figures 3a,c are for accretion velocity, while figures 3b,d are for wind velocity. Panels (a) and (b) are for JS-3/2 DM case, while panels (c) and (d) are for NFW case. In all the figures solid ({\it red online}), long-dashed ({\it blue online}), and 
short-dashed ({\it green online}) curves are for $\Upsilon_B \equiv (33, 100, 390)$. 
To exemplify, here we choose a typical value of $T \sim 6.5 \times 10^6 \, K$, as that used in the context of Fig. 2. 
 }
\label{Fig3}
\end{figure*}

With the decrease in the temperature as $T \lesssim T_{\rm max}$, we enter the domain of multi-transonicity (see Fig. 1). Initially, for higher values of $T$ in the parameter space of 
multi-transonicity, the flow topology is `$X-\alpha$' type. However as the temperature $T$ is further decreased (within the multi-transonic parameter regime), the flow topology changes from `$X-\alpha$' to `$\alpha-X$' type. With even further decrease in the temperature, as $T \lesssim T_{\rm min}$ (see Fig. 1), one eventually leaves the multi-transonic domain, where again the flow exhibits 
uni-transonic behavior. We elucidate the nature of these flow topologies as we proceed through this section. For clarity, in Table \ref{1}, we depict the range in the values of temperature ($T$) for `$X-\alpha$' and `$\alpha-X$' topologies, corresponding to few $\Upsilon_B$ values in the range $\simeq (14-390)$.


\begin{table}
\centerline{{\bf Table 1:} Temperature range for `X-$\alpha$' and `$\alpha$-X' topologies.}
\begin{center} 
\begin{tabular}{|c|c|c|c|c|c|c}
\hline
$\Upsilon_B \, \left({M_\odot}/{L_\odot} \right)$ & $T$ \, ($K$) &  $T$ \, ($K$)  \\  [1ex]
$ $  &   `X-$\alpha$'  &  `$\alpha$-X'  \\  [1ex]
\hline
390 (JS-3/2)  & $\sim (3.58 - 2.825) \times 10^6 $  & $\sim (2.825 - 1.389) \times 10^6$   \\  [1ex]
\hline
390 (NFW)  & $\sim (3.89 - 2.65) \times 10^6 $  & $\sim (2.65 -1.246) \times 10^6$ \\  [1ex]
\hline
100 (JS-3/2)  & $\sim (3.153 - 2.282) \times 10^6 $  & $\sim (2.282 - 1.32) \times 10^6 $     \\  [1ex]
\hline
100 (NFW) & $\sim (2.91 - 2.13) \times 10^6 $  & $\sim (2.13 - 1.243 ) \times 10^6 $    \\  [1ex]
\hline
33 (JS-3/2) & $\sim (2.89 - 2.09) \times 10^6 $   & $\sim (2.09 - 1.274) \times 10^6  $  \\  [1ex]
\hline
33 (NFW) & $\sim (2.8 - 2.03) \times 10^6 $  & $\sim (2.03 - 1.24) \times 10^6  $ 
  \\  [1ex]
\hline
14 & $\sim (2.71 - 1.97) \times 10^6 $  & $\sim (1.97 - 1.235) \times 10^6 $ 
  \\  [1ex]
\hline
\end{tabular}
\end{center}
Note that in column 2, the values of $T$ at the left parenthesis actually represent the corresponding values of $T_{\rm max}$, while in column 3, the values of $T$ at the right parenthesis actually represent the corresponding values of $T_{\rm min}$ (see section III).
\end{table}


\begin{figure*}
\centering
\includegraphics[width=170mm]{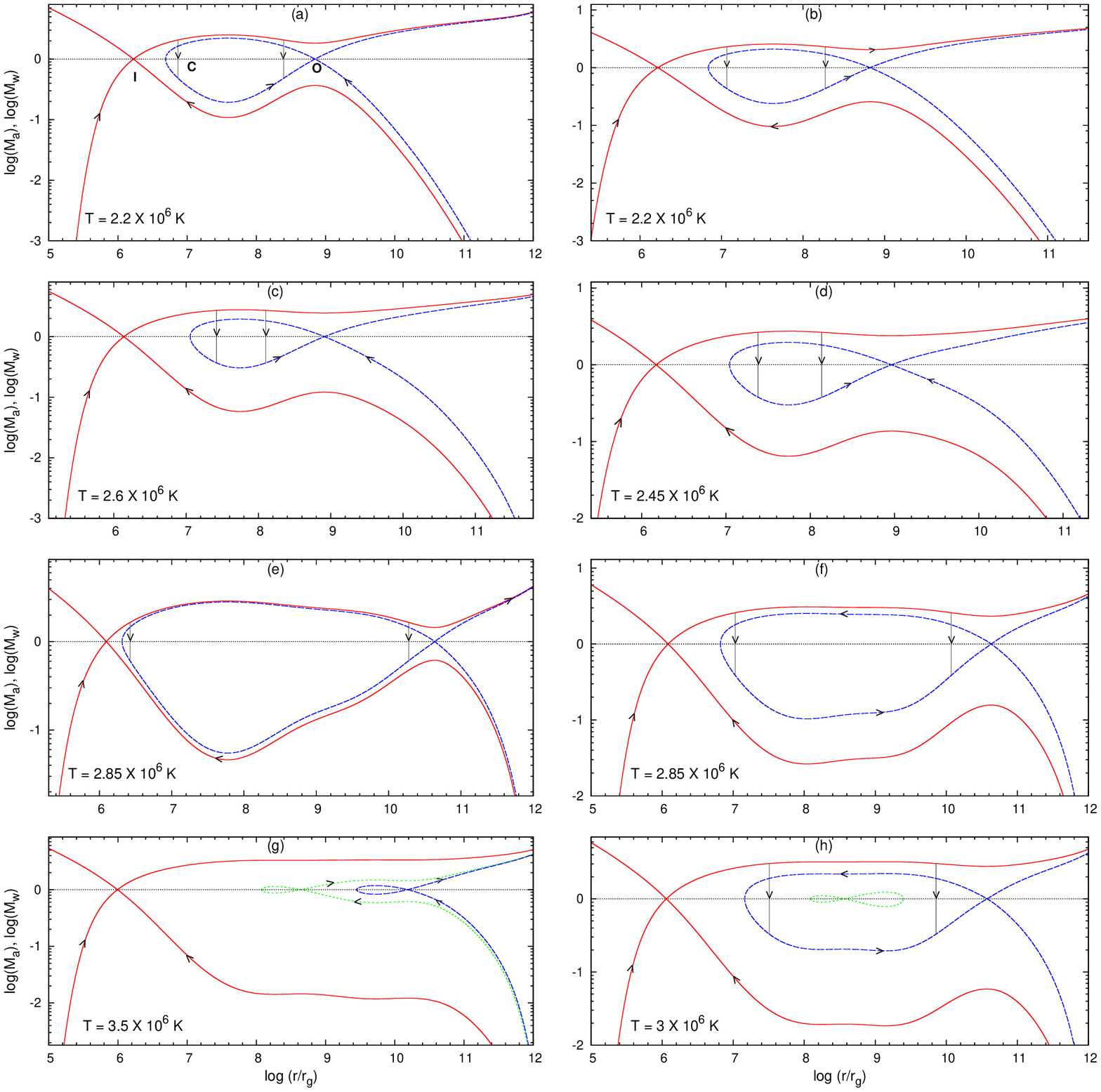}
\caption{Radial profile of Mach number ($M$) representing `$X-\alpha$' type topologies for multi-transonic flows, depicting few sample cases of isothermal shock transitions in spherical wind flows. Figures 4a,c,e correspond to $\Upsilon_B \equiv (33, 100, 390)$, respectively, generated for JS-3/2 DM profile. Figures 4b,d,f resemble figures 4a,c,e, but generated for NFW DM 
profile. Figures 4g,h depict sample cases for five sonic points corresponding to $\Upsilon_B = 390$; Fig. 4g corresponds to JS-3/2 DM case, whereas Fig. 4h corresponds to NFW DM 
case. The corresponding typical values of temperature ($T$) for which the profiles are obtained are given in the respective figures. In all the figures solid ({\it red online}) curves represent the accretion/wind flow solutions through the inner `X-type' sonic point, while the 
long-dashed ({\it blue online}) curves represent the corresponding solutions through the outer `X-type' sonic point. In figures 4g,h (for five sonic point case), the corresponding 
solutions through the middle `X-type' sonic point are represented by short-dashed ({\it green online}) curves. In all the figures, horizontal dotted line corresponds to $M=1$. For clarity, as an example, inner and outer `X-type' sonic points are marked by `$I$' and `$O$' in Fig. 4a, respectively, while the center-type sonic point by `$C$'. The vertical lines marked by arrowheads in all the figures designate the shock transitions in the respective wind solutions. Note that, among the two isothermal shock transitions, only the outer shock transition is stable. 
 } 
\label{Fig4}
\end{figure*}

\begin{figure}
\centering
\includegraphics[width=90mm]{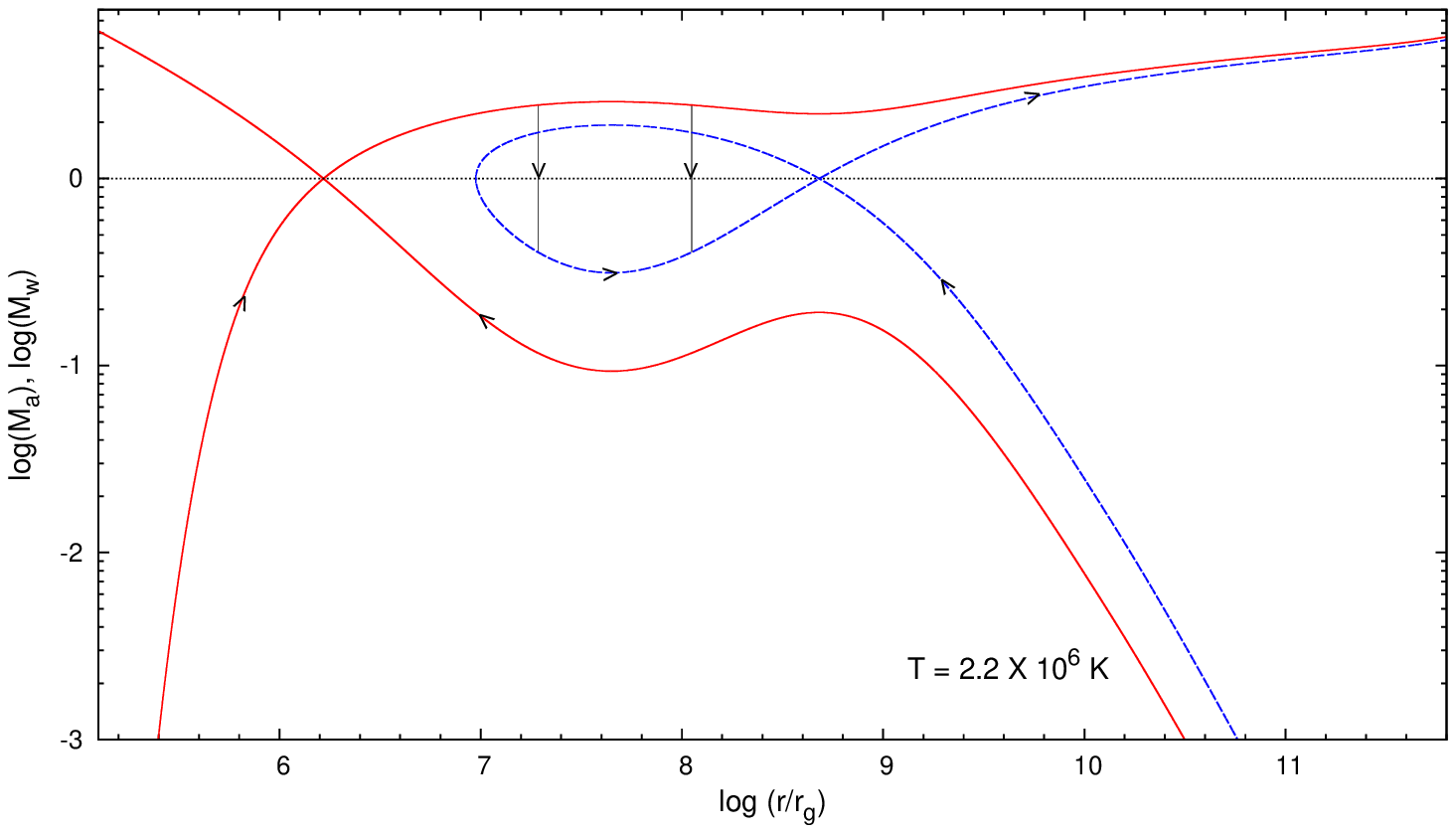}
\caption{Similar to that of Fig. 4, but for $\Upsilon_B = 14$ with no DM scenario. Isothermal shock transitions in wind flows are depicted by vertical lines marked by arrowheads. The corresponding typical choice of temperature for which the profile is generated is given in the respective figure. Here too, note that, among the two isothermal shock transitions, only the outer shock transition is stable.  
 } 
\label{Fig5}
\end{figure}

\subsection{Solutions containing shocks}
\label{4.1}

Shocks are only possible in the multi-transonic domain comprising of at least three sonic points in the 
flow; inner and outer saddle-type (or `X-type') sonic points, and in between, center-type (or `O-type') sonic point. To begin with we concentrate on `$X-\alpha$' type topology. In 
figures 4 and 5, we depict the radial distribution of the Mach number (representing global flow topologies) for few sample cases, for the range of $\Upsilon_B \simeq (14-390)$. Let us 
focus on any of the figures having three sonic points [panel (a-f) in Fig. 4 or Fig. 5]. For three sonic points scenario, inner and outer sonic points are two locations (marked by `$I$' and `$O$' in 
Fig. 4a) through which physical flow is possible, whereas, through the middle one i.e., the O-Type sonic point (marked by `$C$' in Fig. 4a), physical flow is not 
possible. It is being clearly seen from these figures that the inner sonic flow topology represents `$X$' type, whereas through the outer sonic point, the flow topology represents `$\alpha$' type, and hence one refers to them as `$X-\alpha$' topology (for more details see e.g., Abramowicz \& Chakrabarti 1990). As mentioned previously, for relatively higher values of $\Upsilon_B$ ($\Upsilon_B > 200$ for NFW case; $\Upsilon_B > 240$ for JS-3/2 case) even five sonic points could also be obtained in the flow (see \S III). To exemplify, in figures 4g and 4h, we have depicted 
sample cases for five sonic points for two different types of flow topologies corresponding to $\Upsilon_B = 390$; one representing for JS-3/2 DM profile (Fig. 4g), the other representing for NFW DM case (Fig. 4h). Here it needs to be pointed out that for five sonic points scenario, corresponding to both JS-3/2 and NFW profiles, one obtains both types of flow topologies as shown in the 
figures 4g and 4h. The kind of flow topology as represented in Fig. 4h is obtained at a lower range of temperature, whereas, the kind of flow topology as represented in Fig. 4g is obtained at a relatively higher range of temperature. In the same spirit as of the other figures 4(a-f), with respect to the innermost and outermost sonic points, one can associate figures 4(g-h) with $X-\alpha$ type topology.    

\begin{figure*}
\centering
\includegraphics[width=170mm]{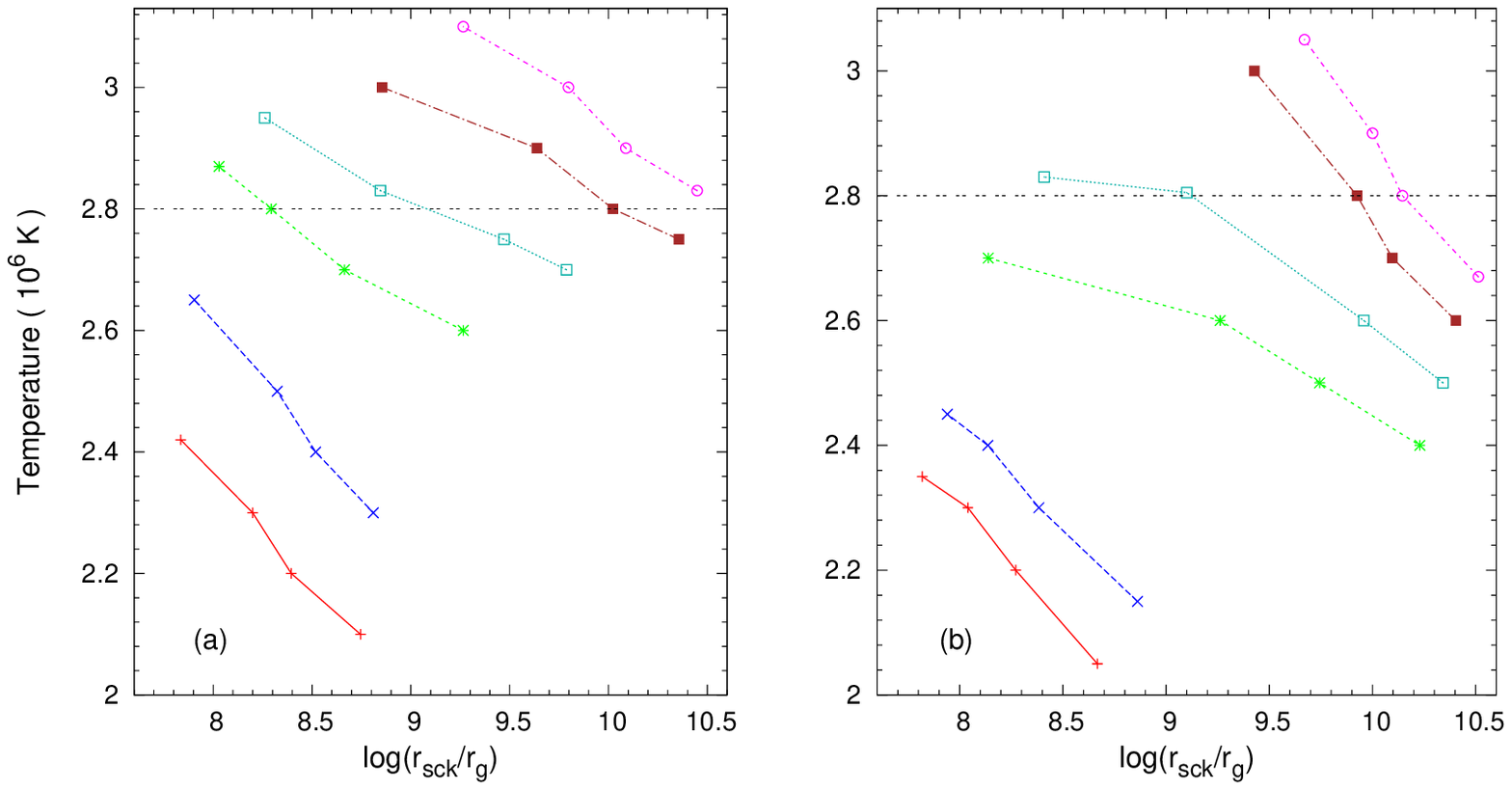}
\caption{Variation of wind shock location ($r_{\rm sck}$) with temperature ($T$) for various values of $\Upsilon_B$. The left panel is for JS-3/2 DM profile, while the right panel is for NFW DM 
profile. The lines from bottom to top in each panel correspond to $\Upsilon_B \equiv (33, 100, 250, 300, 350, 390)$, respectively. In both the panels, horizontal (double-dashed) line corresponds to 
$T = 2.8 \times 10^6\, K$. Note that, $r_{\rm sck}$ corresponds to outer shock location.  
 } 
\label{Fig6}
\end{figure*}

\begin{figure*}
\centering
\includegraphics[width=180mm]{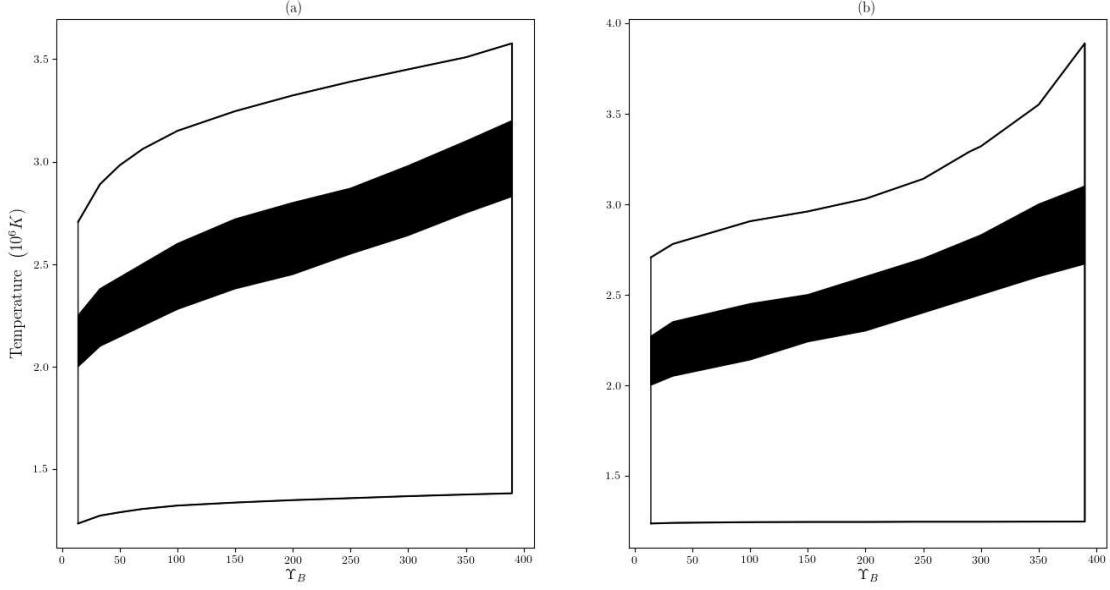}
\caption{Multi-transonic parameter space spanned by $\left < \Upsilon_B, T \right >$ represented by the solid line boundary. Solid filled space depicts the actual shock parameter space in the context of our wind-type flows. The left panel is for JS-3/2 DM case, while the right panel is for NFW DM profile. 
 } 
\label{Fig7}$\Upsilon_B$
\end{figure*}

\begin{figure*}
\centering
\includegraphics[width=180mm]{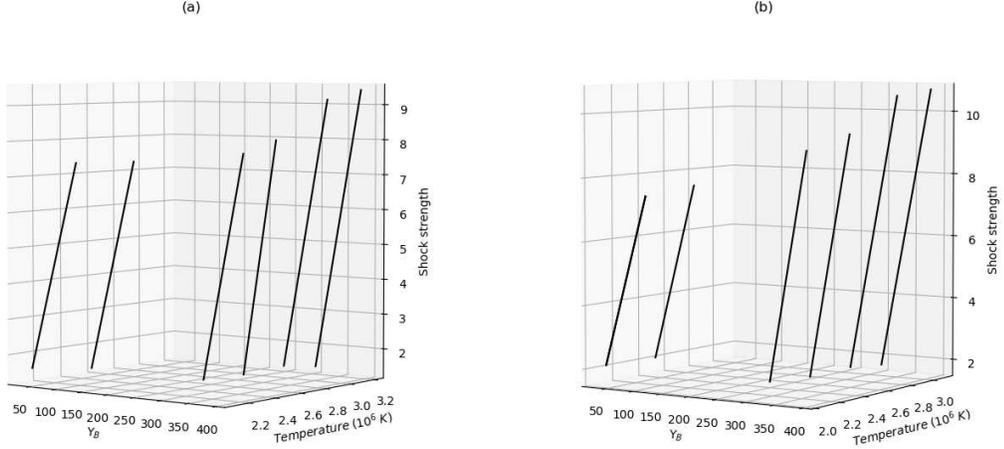}
\caption{ Variation of shock strength with $\Upsilon_B$ and temperature ($T$), corresponding to shock parameter space described in Fig. 7. Shock strength is plotted along $Z$-axis, while $\Upsilon_B$ and $T$ are along $X$ and $Y$ axes, respectively. Figure 8a in the left panel corresponds to JS-3/2 DM profile, while Fig. 8b in the right panel is for NFW DM profile. In both the panels, the solid lines from left to right correspond to $\Upsilon_B \equiv (33, 100, 250, 300, 350, 390)$. 
 } 
\label{Fig8}
\end{figure*}

\begin{figure*}
\centering
\includegraphics[width=180mm]{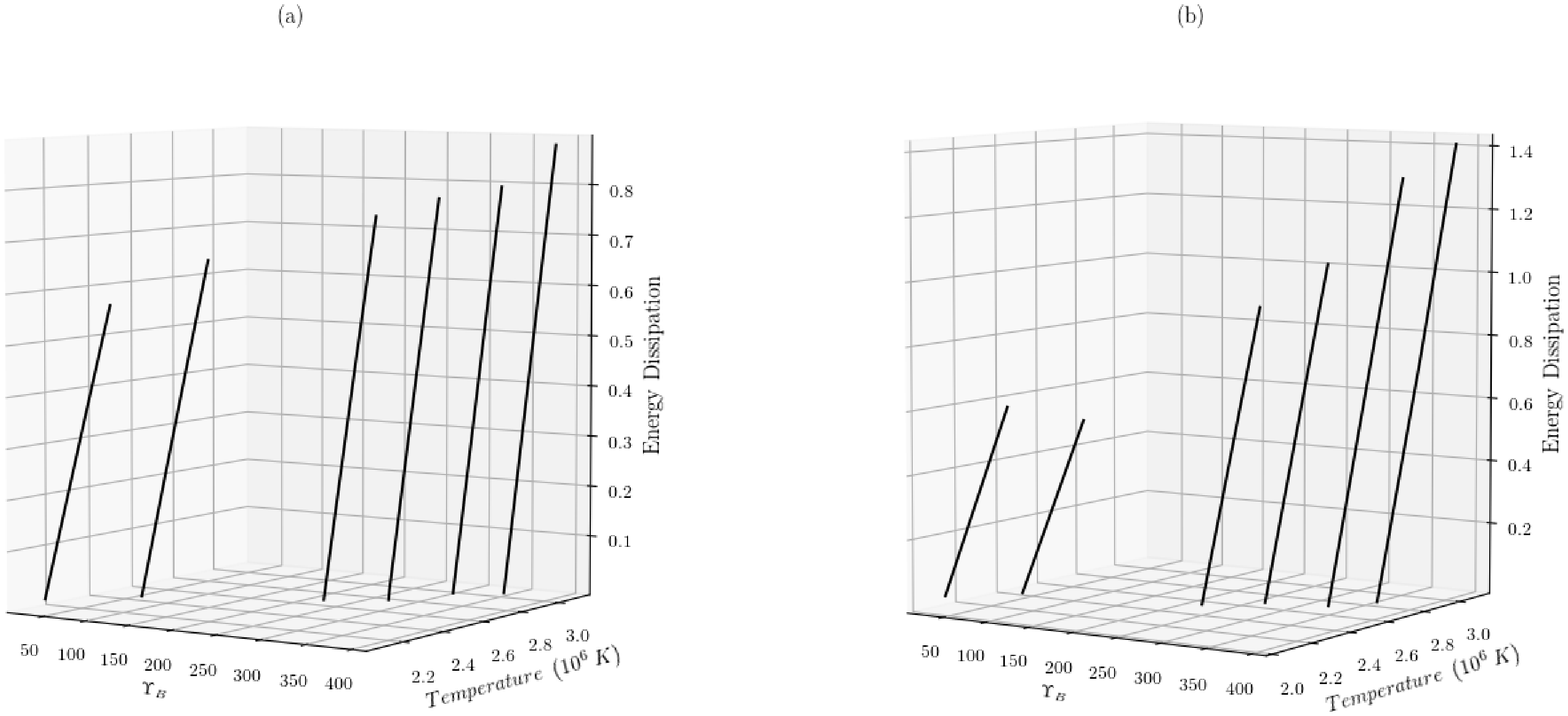}
\caption{ Variation of energy dissipation at shock ($\Delta \mathscr{E}$) with temperature ($T$) and $\Upsilon_B$, corresponding to shock parameter space described in Fig. 7. $\Delta \mathscr{E}$ is plotted along $Z$-axis, while $\Upsilon_B$ and $T$ are along $X$ and $Y$ axes, respectively. Figure 9a in the left panel corresponds to JS-3/2 DM profile, while Fig. 9b in the right panel is for NFW DM profile. In both the panels, the solid lines from left to right correspond to $\Upsilon_B \equiv (33, 100, 250, 300, 350, 390)$ in the units ($10^{-5} c^2$)
 } 
\label{Fig9}
\end{figure*}

Although `$X-\alpha$' type profile is unlikely to generate a shock in accretion branch (see e.g., Abramowicz \& Chakrabarti 1990), nonetheless, it can participate in the possible generation of shocks in wind flow branch, which we are more interested in. We examine the formation of shocks in the flow by checking the isothermal shock conditions (as stated in section III.A) at each radial location throughout the flow spanning the entire multi-transonic parameter space, following the similar numerical technique adopted by earlier authors (e.g., Chakrabarti 1989a; Chakrabarti 1990a; Yang \& Kafatos 1995; Das et al. 2003). Our analysis reveals, that isothermal shocks could be possible in spherical wind-type flows for the entire spectrum of $\Upsilon_B$ corresponding to both NFW and JS-3/2 DM profiles. To exemplify, we choose similar sample cases represented for `$X-\alpha$' topologies in figures \ref{Fig4} and \ref{Fig5}, where we confirmed the shock formation in the wind flow branch. The vertical lines with arrowheads in figures \ref{Fig4} and \ref{Fig5} designate the shock transitions in the respective wind solutions. Here the wind flow initially traverses through the inner X-type sonic point, after which the flow encounters a shock and jumps along the vertical arrowhead lines, eventually traversing outwards through the outer X-type sonic point, with a negative value of $\delta \mathscr{E}$ between $r_{O}$ and $r_{I}$. Note that for five sonic point case, the flow topology depicted in Fig. \ref{Fig4}g is unlikely to permit any shock transitions in wind solutions. 

Resembling the earlier studies in the context of isothermal shocks (e.g., Chakrabarti 1989a; Yang \& Kafatos 1995; Das et al. 2003), here too, our analysis yields two real physical shock locations, represented by two vertical arrowhead lines in the stated figures; one shock location (inner shock) between inner X-type and center-type sonic points, other (outer shock) between center-type and outer X-type sonic point. A pertinent question can then arise as to which of these shocks is stable. Chakrabarti (1989a) performed a local stability analysis to ascertain the stability of these shocks. Later on, Chakrabarti \& Molteni (1993), Yang \& Kafatos (1995), adopted more robust stability analysis to figure out which of these shocks is stable. From their stability analysis Yang \& Kafatos (1995) have pointed out that among these two shocks outer shock is stable and inner shock is unstable, thus obtaining an unique stable shock location. Earlier, Chakrabarti \& Molteni (1993) also found that outer shock is stable, however they did not arrive at any definitive conclusion about the inner shock (Yang \& Kafatos 1995). Following Yang \& Kafatos (1995), Das et al. (2003) in their study concluded that only outer accretion shock is stable. Here, following Yang \& Kafatos (1995), Das et al. (2003), we found that in the context of our wind solutions, outer shock is only stable, while the inner shock is unstable. Hereinafter, in the context of isothermal shocks, we always refer to stable outer shocks. 

Figures \ref{Fig4} and \ref{Fig5} indicate that shock location ($r_{\rm sck}$) is quite sensitive to $\Upsilon_B$, as well as temperature ($T$) of the flow. To ascertain the actual dependency 
of $r_{\rm sck}$ on $\Upsilon_B$ and $T$, in Fig. \ref{Fig6}, the variation $r_{\rm sck}$ with temperature $T$ is shown, for various values of $\Upsilon_B$. Figure \ref{Fig6} reveals that $r_{\rm sck}$ follows a pattern in $\left < \Upsilon_B, T \right > $ parameter space. For a fixed choice of $\Upsilon_B$, with the increase of $T$, $r_{\rm sck}$ moves inward. On the other hand, if $T$ remains 
fixed, with the increase in the value of $\Upsilon_B$, possible shock locations tend to shift outwards. For clarity, a representative line of constant $T = 2.8 \times 10^6\, K$ is marked
(horizontal double-dashed lines in both the panels in Fig. \ref{Fig6}), to highlight this particular trend of $r_{\rm sck}$. Moreover, the figure also reveals that with the increase in the value of 
$\Upsilon_B$, one can obtain possible shocks in the flow at a (relatively) higher temperature range, however for a wider range of $r$, or in other words, shocks can form over a wider radial 
domain. 

In Fig. \ref{Fig7}, we depict the actual shock parameter space spanned by $\left < \Upsilon_B, T \right > $, in the context of our wind-type flows. In the figure, we use the (outer) solid line boundary 
to represent the entire domain of multi-transonicity extending from three sonic points case to five sonic points case comprising of parameter space representing both `$X-\alpha$' and `$\alpha-X$' 
topologies. As stated earlier, it is seen that isothermal shocks can occur in spherical wind-type flows for the entire range of $\Upsilon_B$ $\simeq (14-390)$ considered here, corresponding to 
both NFW and JS-3/2 DM profiles. It is to be noted that shock parameter space depicted in Fig. \ref{Fig7} dwells within the multi-transonic parameter space region that corresponds to only 
`$X-\alpha$' type flow topologies. 

\begin{figure*}
\centering
\includegraphics[width=160mm]{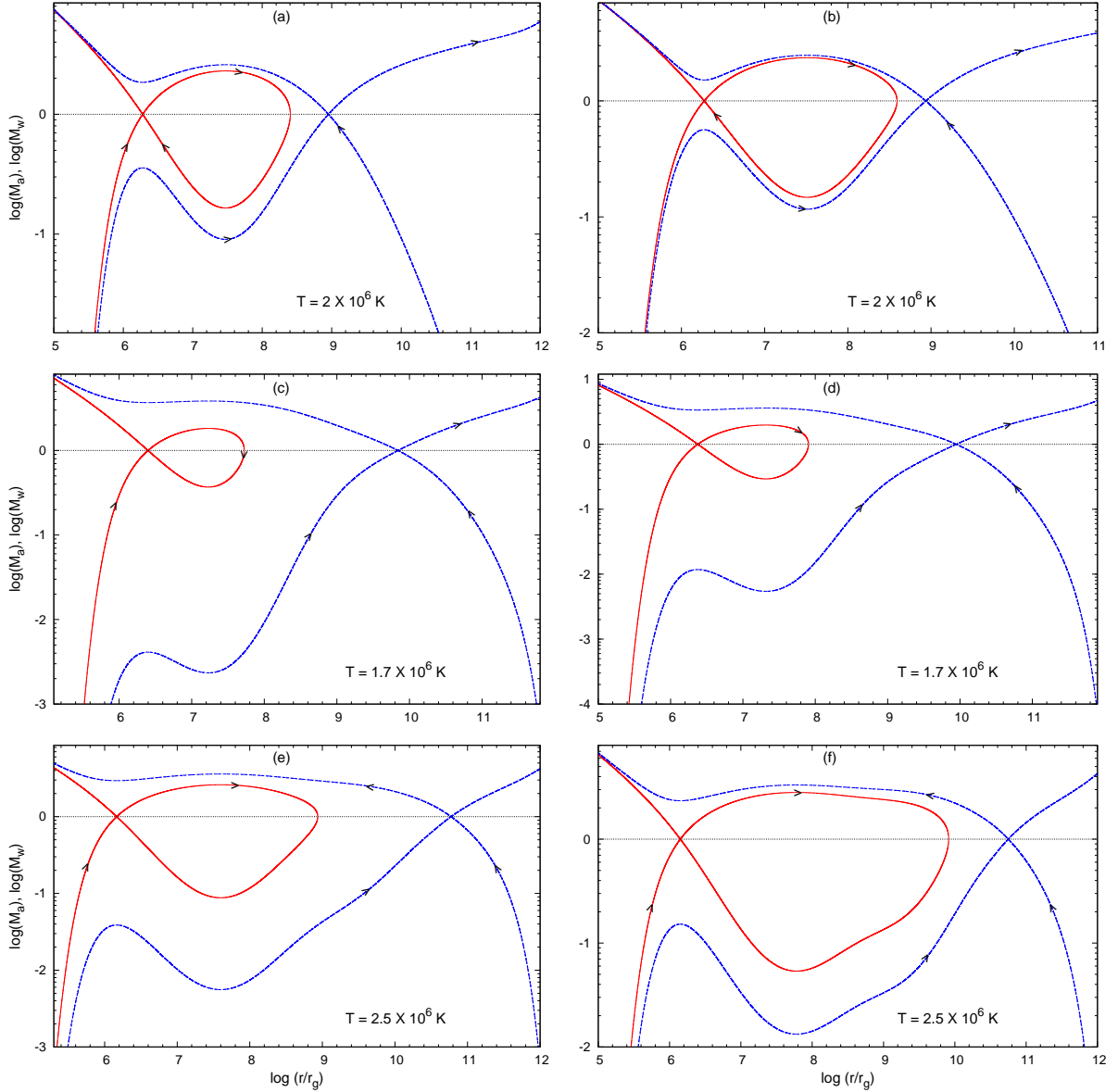}
\caption{Radial profile of Mach number ($M$) representing `$\alpha-X$' type flow topologies for multi-transonic flows, shown for a few sample cases. Figures \ref{Fig10}a,c,e correspond to $\Upsilon_B \equiv (33, 100, 390)$, respectively, generated for JS-3/2 DM profile. Figures \ref{Fig10}b,d,f resemble figures \ref{Fig10}a,c,e, but generated for NFW DM 
profile. In the respective figures, we have shown the typical values of temperature ($T$) to generate our profiles. In all the figures long-dashed ({\it blue online}) curves represent the accretion and wind flow solutions through the outer `X-type' sonic point, while the solid ({\it red online}) curves represent the corresponding solutions through the inner `X-type' sonic point. Horizontal dotted line in all the figures is for $M=1$. Note that, here we do not obtain any isothermal shock transitions.  
 } 
\label{Fig10}
\end{figure*}

\begin{figure*}
\centering
\includegraphics[width=160mm]{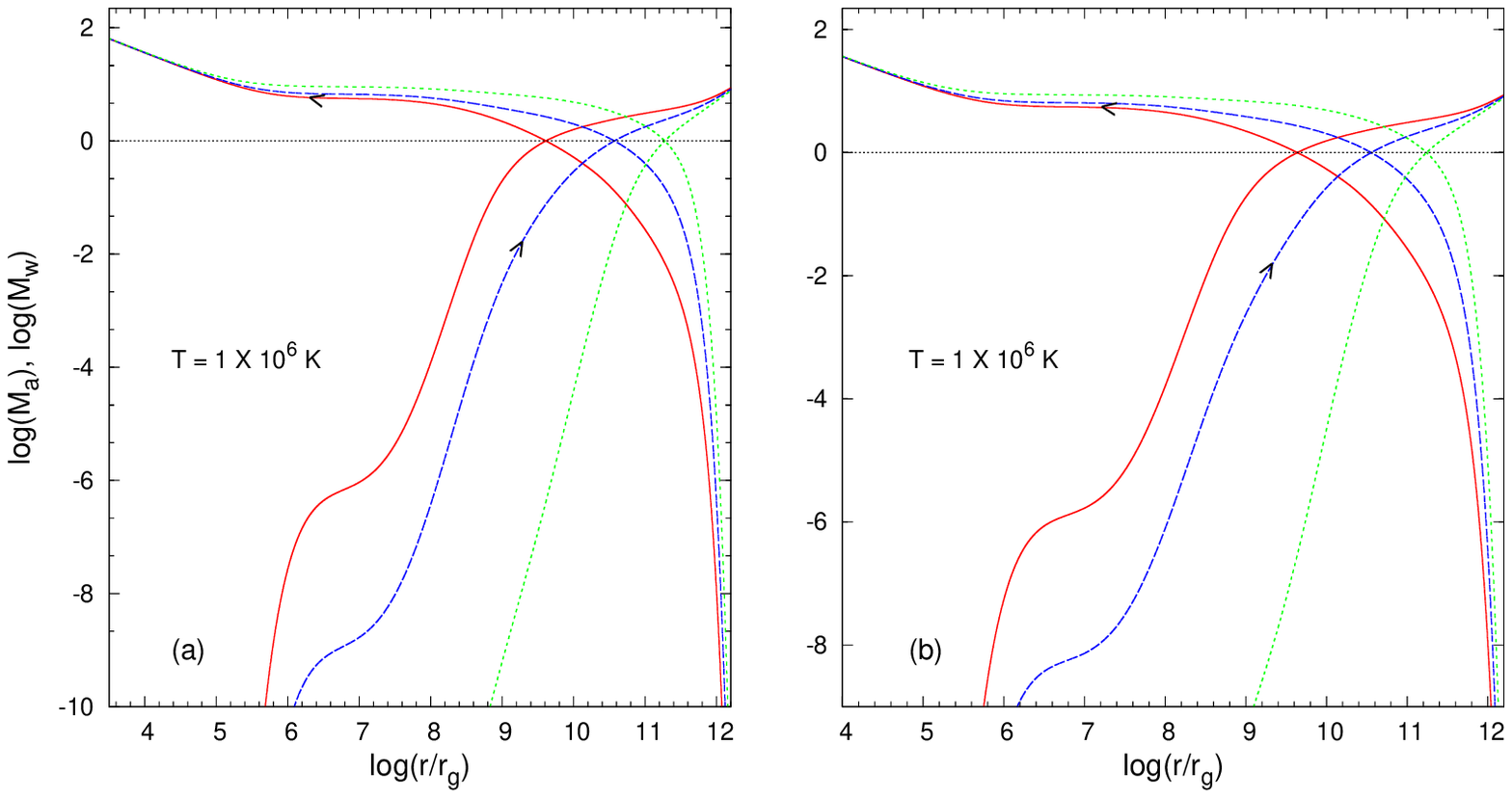}
\caption{Radial profile of Mach number ($M$) for uni-transonic regime where the accretion and wind flow solutions pass only through the outer `X-type' sonic point, shown for a few sample cases. Figure \ref{Fig11}a in the left panel corresponds to JS-3/2 DM profile, while Fig. \ref{Fig11}b in the right panel is for NFW DM profile. In each panel, solid ({\it red online}), long-dashed ({\it blue online}), and short-dashed ({\it green online}) curves represent accretion and wind solutions 
correspond to $\Upsilon_B \equiv (33, 100, 390)$, respectively. Representative arrowheads have been drawn to indicate accretion/wind flow solutions. In the respective figures, we have shown the typical values of temperature ($T$) to generate our profiles. Horizontal dotted line in both panels is for $M=1$. 
 } 
\label{Fig11}
\end{figure*}

In Fig. \ref{Fig8}, we plot the variation of the strength of shock (which is measured by the pre-shock to post-shock Mach number ratio), as a function of $\Upsilon_B$ and $T$ corresponding to shock parameter space described in Fig. \ref{Fig7}. It is to be noted that for isothermal shocks considered here, shock strength is exactly equivalent to the shock compression ratio (e.g., Das et al. 2003). It is seen from the figure that the shock strength increases with the increase in both $T$, as well as $\Upsilon_B$. More interestingly, the strength of the galactic induced shocks that occur in our 
wind-type flows in the central to outer radial locations, are found to be comparable to that of the shocks in the advective flows obtained in the vicinity of the BH; NFW DM profile yields somewhat higher shock strength, even of magnitude $>10$ for some region of parameter space. Isothermal shock transitions are characterized by dissipation of energy at the shock location. In Fig. \ref{Fig9}, we show the variation of the amount of energy dissipated $\Delta \mathscr{E}$ for galactic induced shocks as a function of $T$ and $\Upsilon_B$, corresponding to shock parameter space described in Fig. \ref{Fig7}. The figure shows that resembling the scenario in Fig. \ref{Fig8}, here too, $\Delta \mathscr{E}$ increases with the increase in 
$T$ as well as $\Upsilon_B$, implying that more the strength of the shock (i.e., stronger the shock), greater amount of energy would then be dissipated at the shock location.

It is worthy to note that the shock quantities, such as shock location, shock strength and energy dissipated at the shock, correlate well with galactic mass-to-light ratio $\Upsilon_B$; with the increase in $\Upsilon_B$, not only the shock position $r_{\rm rck}$ shifts outward, but also both the shock strength and $\Delta \mathscr{E}$ get enhanced. This is owing to the fact that with the increase in galactic mass-to-light ratio, the centrifugal potential barrier moves further outwards, as well the strength of the galactic gravitational field gets augmented, and more amount of gravitational potential energy would then be available to get released at the shock location. Our analysis reveals that shock strength and $\Delta \mathscr{E}$ can increase both in the radially outward and inward directions, depending upon $\Upsilon_B$ and $T$, respectively. This is unlike the scenario in case of the shocks in advective flows in the vicinity of BH/central object, where these quantities are found to increase only in the inward direction as one moves closer to the central object (e.g., Chakrabarti 1989a; Das et al. 2003, although studied in the context of accretion shocks).

\begin{table}
\centerline{{\bf Table 2:} Radial range of `forbidden region' for `$\alpha-X$' topology}
\begin{center}
\begin{tabular}{|c|c|c|c|c|c|c|c|c|c|c|c|c|c|c|}
\hline
$\Upsilon_B \, \left({M_\odot}/{L_\odot} \right)$ & Temperature \, ($K$) & Radial range \, ($r_g$)  \\  [1ex] 
$ $ & $ $ & (log value) \\  [1ex]
\hline
390 (JS-3/2) & $ 2.823 \times 10^6$ & $\sim 10.5782 - 10.647$ \\  [1ex] 
&   $ 1.4 \times 10^6$ & $\sim 6.9342 - 11.157$ \\  [1ex]
\hline
390 (NFW) & $ 2.649 \times 10^6$ & $\sim 10.4736 - 10.699$ \\  [1ex] 
& $ 1.28 \times 10^6$ & $\sim 7.1227 - 11.1489$ \\  [1ex] 
\hline
100 (JS-3/2) & $2.281\times 10^6$ & $\sim 9.13588 - 9.1548$ \\  [1ex] 
&  $1.35 \times 10^6$ & $\sim 7.16677 - 10.1889$ \\  [1ex] 
\hline
100 (NFW) & $ 2.125 \times 10^6$ & $\sim 9.15 - 9.31$ \\  [1ex] 
& $ 1.25  \times 10^6 $ & $\sim 6.907 - 10.3498$ \\  [1ex] 
\hline
33 (JS-3/2) & $ 2.084 \times 10^6$ & $\sim 8.7667 - 8.9048$ \\  [1ex] 
&  $ 1.28 \times 10^6$ & $\sim 6.8945 - 9.377$ \\  [1ex] 
\hline
33 (NFW) & $2.0 \times 10^6 $ & $\sim 8.59 - 9.9368$ \\  [1ex] 
& $ 1.26 \times 10^6$  & $\sim 7.151 - 9.377$ \\ [1ex]
\hline 
14 & $ 1.94 \times 10^6$ & $\sim 8.37 - 8.83 $  \\  [1ex] 
& $1.25 \times 10^6 $ & $\sim 7.102 - 9.1269$  \\  [1ex] 
\hline
\end{tabular}
\end{center}
Note that corresponding to each specific case of $\Upsilon_B$ (given in column 1), in column 2 the upper temperature value represents that value of temperature which is near to the corresponding $X-\alpha$ to $\alpha-X$ transition temperature, whereas the lower temperature value represents that value of temperature which is near to the corresponding `lower limit temperature' for $\alpha-X$ topology ($T_{\rm min}$) (for clarity see Table \ref{1}). Note that for $T$ $\lesssim T_{\rm min}$, the flow topology changes from $\alpha-X$ to that illustrated in Fig. \ref{Fig11}. 
\end{table}

\subsection{Solutions without shocks}
\label{4.2}

In the previous subsection we concentrate on the multi-transonic parameter space regime representing `$X-\alpha$' type topology, where one have the possibilities to obtain shocks 
in the wind-type flow. With the decrease in the flow temperature (within the domain of multi-transonicity), the flow topology changes from `$X-\alpha$' to `$\alpha-X$' type (see the transition temperature values in Table \ref{1}). In Fig. \ref{Fig10}, we depict the radial distribution of the Mach number for a few sample cases, representing `$\alpha-X$' topologies. It is being clearly seen from the figure that the outer sonic flow topology represents `$X$' type through which the flow connects infinity and the BH event horizon, whereas the flow topology through the inner sonic point represents `$\alpha$' type (see e.g., Abramowicz \& Chakrabarti 1990). Here the `$\alpha$' type topology implies that although the wind flow begins its journey traversing through the inner X-type sonic point, the matter eventually does not find any possible physical path to proceed further outwards, following the accretion branch it then enters the BH horizon. Although, `$\alpha-X$' type profile is unlikely to participate in the possible generation of shocks in the wind solution, they are more likely to generate a shock in the accretion flow branch. Nonetheless, in our case, we do not obtain any isothermal shocks in accretion branch. 

Finally, if one decreases the temperature further more as $T$ $\lesssim T_{\rm min}$, multi-transonic nature of the flow disappears and the flow displays uni-transonic behavior. However, unlike the other uni-transonic regime, where the inner `X-type' sonic point only provides the possible physical path for the flow to make a sonic transition, in this case, it is the outer `X-type' sonic point that provides the only possible physical path for the flow to make a sonic transition. The corresponding flow topology is shown in Fig. \ref{Fig11}, for a few sample cases. 

It is evident from the Mach number profiles in Fig. \ref{Fig10}, that in the context of `$\alpha-X$' topology, any accretion flow that begins subsonically whose outer accretion boundary 
radius (or ambient radius $r_{\rm out}$) lies outside the `$\alpha$' loop and within the outer X-type sonic point ($r_{O}$), the flow would not be able to make any sonic transition to make its way onto the central BH. This then implies that, corresponding to those temperatures for which the flow topology represents `$\alpha-X$' type, there is some sort of `forbidden 
radial region' in the ambient medium, from within which no physically realistic Bondi-type accretion could possibly be triggered. To comprehend about the radial extent of this 
`forbidden region', in Table \ref{2}, corresponding to a few temperature values, we present the radial range of this `forbidden region' for few $\Upsilon_B$ values in the range $\simeq (14-390)$. It is seen from Table \ref{2}, that with the decrease in the temperature, the radial extent of this `forbidden region' gets augmented. A similar scenario also arises for the case illustrated in 
Fig. \ref{Fig11}, however, there the `forbidden region' corresponds to the entire radial region inside the outer X-type sonic point. 


\section{Mass inflow rate}
\label{5} 

Bondi mass inflow rate is quite a relevant quantity in the context of LERGs powered by hot mode accretion. Even though a pure spherical Bondi-type accretion may seem to be far from reality, however, in the context of massive ellipticals, Bondi prescription is typically adopted to estimate the requisite mass supply rate onto their host active nucleus (at least in the first approximation), and could be quite useful to quantitatively model the radio-mode (or maintenance-mode) feedback (see \S I and references therein). In fact, earlier investigations show that in these LERGs, Bondi accretion rate is tightly correlated with jet kinetic power (e.g., Allen et al. 2006; Nemmen et al. 2007; Narayan \& Fabian 2011). Nonetheless, in this context, one usually employs the (simple) 
classical Bondi solution (Bondi 1952) to quantify the required mass supply rate (e.g., Hardcastle et al. 2007; Heckman \& Best 2014, and references therein). As noted in \S I, many previous studies have shown that there may be plausible scenarios, when mass inflow rate can be suppressed. In this context, it is worthy to note that in RGJ18, the authors interestingly found that owing to the impact of galactic gravitational potential, Bondi mass inflow rate tends to get {\it augmented} for adiabatic class of flows. Before embarking on to investigate such effect of galactic potential on more realistic accretion scenario, it would be tempting to examine whether such an influence of galactic potential on mass inflow rate remains consistent for other class of polytropic flows. Here, we investigate such an effect for our isothermal case of present interest. In the steady state, in terms of transonic variables, one can define Bondi mass inflow rate ($\dot M_B$), which follows 

\begin{eqnarray}
\vert \dot M_{B} \vert = 4 \pi r^2_c \, c_{\rm sc} \, \rho_c \, ,   
\label{12}
\end{eqnarray}
For isothermal flows pertaining to our five-component galaxy model, using Eqn. (9), 
one can evaluate $\rho_c$ (which is the density at $r_c$) in terms of $\rho_{\rm out}$ (which is the density of the ambient medium), given by 

\begin{eqnarray}
\rho_c = \rho_{\rm out} \exp \left[ \frac{1}{c^2_{\rm sc}} \,\left( \Psi_{\rm Gal} (r_{\rm out}) - \Psi_{\rm Gal} (r_c) - \frac{u_{c}^2}{2} + \frac{u_{\rm out}^2}{2} \right) \right]  \, , 
\label{13}
\end{eqnarray}
where $u_{\rm out}$ is the velocity at outer accretion boundary radius (or ambient radius) $r_{\rm out}$. 
Using Eqns. (4) and (13), one can then express $\dot M_{B}$ through the following relation 

\begin{eqnarray}
\vert \dot M_{B} \vert = 2^{3/2} \pi r^{5/2}_c [\mathscr{F}_{\rm Gal} (r_c)]^{1/2} \rho_{\rm out} \, \,  \exp \left[\frac{1}{c^2_{\rm sc}} \, \left( \Psi_{\rm Gal} (r_{\rm out}) - \Psi_{\rm Gal} (r_c) - \frac{u_{c}^2}{2} +  \frac{u_{\rm out}^2}{2} \right) \right] \, 
\label{14}
\end{eqnarray}
For classical Bondi accretion onto a non-rotating BH in the Paczy\'nski \& Witta (1980) Pseudo-Newtonian regime, 
corresponding Bondi accretion rate is given by 
\begin{align}
\vert \dot M_{B} ({\rm BH}) \vert \approx 2^{3/2} \pi \frac{r^{5/2}_c}{r_c -2} \, \rho_{\rm out} 
\exp \left[\frac{1}{c^2_{\rm sc}} \, \frac{3 r_c -8}{4 (r_c-2)^2} \right] \, ,
\label{15}
\end{align}
where, $r_c$ [in Eqn. (15)] corresponds to the classical Bondi case.  

\begin{figure*}
\centering
\includegraphics[width=170mm]{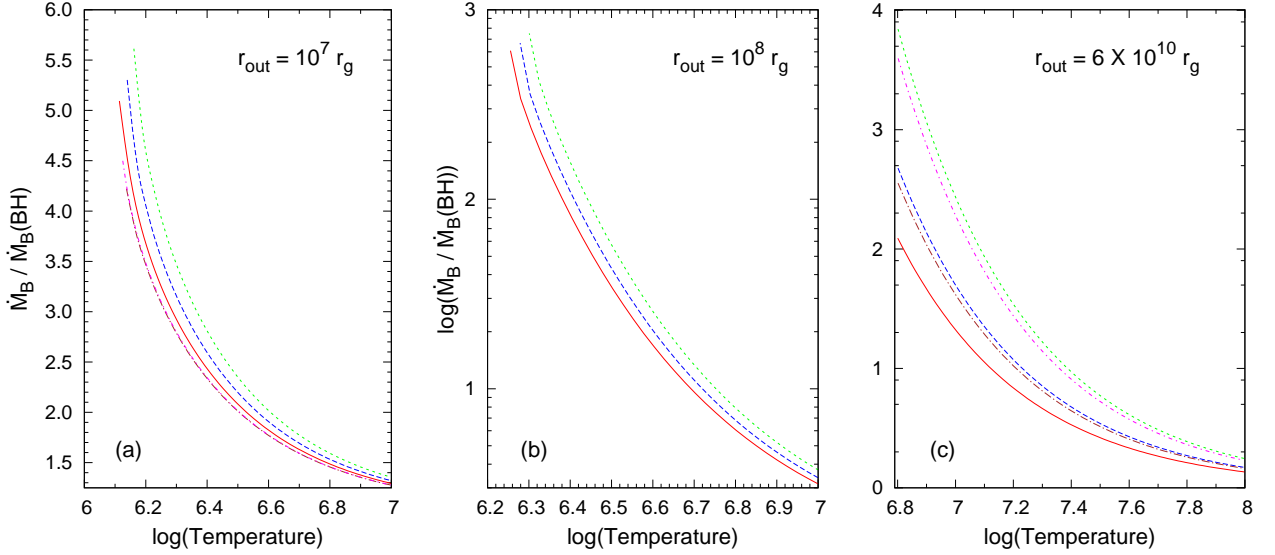}
\caption{Variation of $\frac{\dot M_{B}}{\dot M_{B} ({\rm BH})}$ as a function of ambient temperature $T_{\rm out}$ for different outer accretion boundary radius $r_{\rm out}$, for few sample cases. In figures \ref{Fig12}a,b the profiles are generated with $r_{\rm out} \simeq (10^7 \, r_g, 10^8 \, r_g)$, respectively, approximately resembling an
accretion from hot ISM at the center of a galaxy; whereas in Fig. \ref{Fig12}c the profiles are generated with $r_{\rm out} \simeq 6 \times 10^{10} \, r_g$ corresponding to the flows from ICM/IGM. In all the figures solid ({\it red online}), long-dashed ({\it blue online}), short-dashed ({\it green online} correspond to $\Upsilon_B \equiv (33, 100, 390)$ for JS-3/2 DM profile. For comparison, in Figures \ref{Fig12}a,c we show the corresponding profiles for NFW DM case for $\Upsilon_B = 100$ [long dotted-dashed ({\it brown online})], and $\Upsilon_B = 390$ [short dotted-dashed ({\it pink online})]. Note that in Fig. \ref{Fig12}a, long dotted-dashed and short dotted-dashed overlaps.  
 } 
\label{Fig12}
\end{figure*}

Figure 12 depicts the variation of $\frac{\dot M_{B}}{\dot M_{B} ({\rm BH})}$ as a function of ambient temperature, for different values of ambient radius $r_{\rm out}$. 
Figures \ref{Fig12}a,b approximately resemble the scenario for an accretion flow triggered from the hot interstellar
medium (ISM) phase at the center of a galaxy (e.g., Narayan \& Fabian 2011; RGJ18), while Fig. \ref{Fig12}c approximately corresponds to the scenario for an accretion flow triggered from Intracluster medium (ICM)/intergalactic medium (IGM) (for details see RGJ18 and references therein). 
Figure \ref{Fig12} reveals that owing to the effect of galactic potential Bondi mass inflow rate tends to get enhanced, for same cases, even with orders of magnitude higher relative to the classical Bondi scenario; for elliptical galaxies with higher galactic mass-to-light ratios, as well as for flows having larger accretion outer 
boundary radius (or larger ambient radius), there is a greater enhancement of mass inflow rate. Such augmentation of Bondi accretion rate may have interesting ramifications in the context of mass supply rate required to power the host nucleus of these massive ellipticals. We briefly comment on this aspect in the next section. It is to be noted that in 
figures \ref{Fig12}a,b, curves get truncated corresponding to certain values of $T$ (say $T_{\rm tr}$). The truncation of these curves imply that, those values of temperature $T$ that are less than $T_{\rm tr}$ are associated with the `forbidden region' (see section IV.B), from where Bondi accretion is unlikely to be triggered. 


\section{Discussion and concluding remarks}
\label{6} 

The principal purpose of the present paper is to comprehensively examine the issue of galactic potential induced shock formation in spherical/quasi-spherical Bondi-type flows in the context of low excitation radio galaxies, the more appealing, (dissipative) isothermal standing shocks. In this context, we extended the previous work of RGJ18 to isothermal spherical flows, and also to check whether the findings in our previous study (RGJ18) remain consistent for other class of polytropic flows, like isothermal flows. 
Here one should note that, in the context of accretion related phenomena most of the studies related to issue of shocks predominantly revolves around (dissipationless) Rankine-Hugoniot type shocks for adiabatic class of flows, unlike that for dissipative shocks in isothermal flows. This, perhaps, could be owing to fact that the global isothermality in the flow is hard to preserve. Nonetheless, one should remember that for Rankine-Hugoniot type shock scenarios, the flow structure changes abruptly at the shock, and it is virtually impossible to furnish the accurate pressure balance condition at the shock location {\footnote{Although this is more applicable in the context of flows which are not strictly spherical.}}. On the contrary, for isothermal flows with dissipative shocks, the regular flow structure is maintained through the shock, and one can obtain a correct pressure balance condition at the shock location. Moreover, it has been pointed out that isothermal shocks provide the most effective mechanism to release energy from the flow. Here we like to point out that several studies based on stellar kinematics, kinematics of planetary nebulae, analysis of X-ray data, strong lensing, combination of strong lensing and stellar kinematics, joint strong- and weak lensing survey, have strongly suggested that total gravitational potential of early-type elliptical galaxies tends to be approximately isothermal for a wide range of scales, even though the individual mass profiles may deviate from being isothermal (for details see e.g., Nulsen 1998; Koopmans et al. 2006; Gavazzi et al. 2007; van de Ven et al. 2009; Churazov et al. 2010; and references in them). For Bondi-type spherical/quasi-spherical hot-mode accretion, since the temperature of the gas approximately `follows' the gravitational potential (in the sense that gas is approximately locally virialized), in the main body of the galaxy where the galaxy potential likely dominates (the region of our interest in the present study), roughly isothermal gas would then tend to exist in such regions of the flow. Although, deviations from isothermality are certainly present at some level, in the inner and outer radii. 
A critical analysis of isothermal flows, particularly in the context of dissipative shocks, has quite a relevance in the present context. 

Our present study demonstrates that owing to the influence of galactic potential of the host elliptical, the flow topology gets significantly altered from that of the classical Bondi solution. In the global parameter space spanned by $\left < \Upsilon_B, T \right >$, four distinct types of topologies emerge in the context of our spherical flow: in the uni-transonic regime, the flow topologies are of `$X$' types as illustrated in in figures \ref{Fig2} and \ref{Fig11}; in the multi-transonic parameter space domain, the corresponding flow topologies represent `$X-\alpha$' and `$\alpha-X$' types (figures \ref{Fig4}-\ref{Fig5}, Fig. \ref{Fig10}). This in fact is quite intriguing, in the sense, that such appearances of `$X-\alpha$' or `$\alpha-X$' topologies precisely resemble the scenario in advective flows in the vicinity of BHs/compact objects. It then seems that the galactic potential (represented uniquely through the parameter $\Upsilon_B$) plays an equivalent role in the central to the outer radial regions of our spherical flow, as that of the angular momentum $\lambda$ in the inner regions of the advective flow. Although in our study we do not find shocks in the accretion flow branch, however, on account of the impact of galactic potential, standing isothermal shocks do emerge in the central to the outer radial locations of spherical wind-type outflows (beyond kiloparsec-scale) for the entire range of $\Upsilon_B$, corresponding to both NFW and JS-3/2 DM profiles (i.e., in both the limits of inner slope $\gamma$). The strength of these galactic induced shocks in wind-type outflows are found to be comparable to that of the shocks one would obtain in the advective flows in the vicinity of the BH, even of magnitude $\gtrsim 10$ for some region of parameter space (Fig. \ref{Fig8}). Furthermore, our analysis reveals that shock parameters remain sensitive to $\Upsilon_B$; with the increase in $\Upsilon_B$, not only the shock location moves further outward, but also both shock strength and the amount of energy dissipated at the shock get enhanced. On previous occasions many authors have discussed the possibilities of having multi-transonicity and shocks in wind-type outflowing solutions, either in the context of black hole accretion or in the context of solar and stellar winds (for details see the fourth paragraph in \S I). Here we would like to point out that in large fractions of AGNs, particularly in luminous AGNs and Quasars, wide-angle wind-type outflows from active nucleus are commonly observed around galactic centers, usually called disk-winds. These disk-winds are thought to interact with the surrounding interstellar gas and produce shocks around galactic centers on scales of about kiloparsec (see e.g., Bland-Hawthorn et al. 2007; Tadhunter 2008; King \& Pounds 2015; Tombesi 2017; Morganti 2017). Here it needs to be noted that, galactic potential induced shocks that we obtained in our spherical wind-type outflowing solutions appears to be relevant at much larger scales, beyond kiloparsecs. 

Occurrence of galactic potential induced shocks in spherical `wind-type' flows may have interesting ramifications, in the context of realistic outflows and jets. Without going into the detailed model(s) on formation, structure and evolution of realistic outflows/jets, in a simplistic approximation, one can envisage spherical wind-type flow 
as some form of spherical outflow of matter. In realistic outflows and jets, various features such as hot spots, knots and flaring are believed to be associated with the shocks in the flow. The conventional picture of shock formation in astrophysical outflows/jets usually rests on two plausible scenarios: (i) internal shocks, and (ii) shocks arising via the mediation of the external ambient medium at a relatively large distance from the central object. Radio-emitting jets in AGNs appear in a variety of length and terminal speeds which one usually relates to the impact of ambient medium on jet flow (see Vyas \& Chattopadhyay 2017 and references therein). In low excitation radio galaxies which are usually hosted by massive ellipticals (that we focused in our study) the radio jets emanating from the active nucleus show two distinct morphologies classified as Fanaroff \& Riley (FR) dichotomy (FRI and FRII types). The morphological dichotomy seems to reflect the way the jet/outflowing matter can propagate through the host galaxy and the method of energy transport in the radio source. FRII jets remain largely relativistic and collimated out to the end of radio-lobes often out to scales of hundreds of kiloparsec or more with well-defined jet termination shocks and appear to be able to transport energy efficiently to the end of the lobes. On the other hand, FRI sources are known to be decelerating to sub-relativistic speeds at a relatively lesser distance from the active nucleus (although beyond kiloparsec-scale, over scales of few tens of kpc), are inefficient in transporting energy, less collimated, and are more distorted and plume like (e.g., Laing \& Bridle 2002, also see Hardcastle 2008 for more details on this aspect). It is being commonly perceived that this morphological dichotomy arises owing to either a) the influence of the external ambient medium on the jet structure, or b) due to inherent differences in the jet production mechanism (e.g., Turner \& Shabala 2015). In this context, one can raise a very crucial question: whether, resembling the scenario in spherical `wind-type' outflowing solutions, could galactic potential induced shocks be possible to occur in realistic outflows and jets? Such mode of shock formation (if indeed possible in realistic outflows and jets) would have two major attributes: (i) they would more likely to occur in central to the outer regions of the outflow beyond kiloparsec-scale, and depending on galactic mass-to-light ratio (which may be different for different elliptical galaxies) such shocks may form at different radial locations for different ellipticals (see \S IV.A), and (ii) these shocks are terminal type shocks that could cause sudden deceleration of outflowing matter, and thus could substantially affect the outflow/jet speed. Such galactic potential induced shocks (if possible in realistic outflows/jets) could then considerably affect the dynamics and the propagation of outflows/jets in the medium beyond kiloparsec-scale, and the impact of these shocks on outflow/jet structure would expected to be similar as that of the other two scenarios discussed above (points a, b). In this context, galactic potential shocks could be envisaged as one of the `potential contributing factors' to account for the observed morphological dichotomy in the radio sources. Although, here one can only make a speculative guess, and whether such a scenario could be feasible in practice is hard to predict at the current juncture. At the least, it would be worthy to consider a simple jet flow model (like a conical shape outflow model) as being adopted in the previous works of Vyas \& Chattopadhyay (2017) or Ferrari et al. (1985), and examine the possibilities of the formation of galactic induced shocks in such collimated type outflowing solutions. Such a task, although is beyond the scope of this study, shall be undertaken in the foreseeable future. 

Isothermal shock transitions are characterized by the dissipation of energy at the shock location, which might have interesting consequences in the context of realistic outflows/jets. Although in our analysis we found that the amount of energy dissipated at the shock in the context of our wind-type outflowing solutions is a small fraction of the total rest mass energy of the flow [maximum being of the order of $\sim 10^{-3} \, \%$ for some region of parameter space (Fig. \ref{Fig9})], however, this may not be inconsequential in the context of realistic outflows/jets, provided, such type of galactic potential shocks actually occur in realistic outflows/jets. To exemplify, we roughly consider a typical case of M87 jet: Here, for our purpose, we simply presume for the time being that the amount of specific energy dissipated at the shock ($\Delta \mathscr{E}$) in the context of our wind-type flow, is roughly of the similar magnitude one would expect if such isothermal shocks occur in realistic outflows/jets, which may not be an unreasonable approximation. If ${\dot M_j}$ is the mass outflow rate or the mass loss rate into the jet, then one can roughly say that the amount of power associated with this dissipation is $\approx {\dot M_j} \, \Delta \mathscr{E}$. For M87, which is a nearby giant elliptical galaxy, we do have some reliable estimate of total jet kinetic power whose magnitude is $\approx 10^{44} \, {\rm ergs \, s^{-1}}$ (e.g., Bicknell \& Begelman 1996; Forman et al. 2005). However the core radio luminosity is inferred to be $\approx 10^{39.8}\, {\rm ergs \,s^{-1}}$ (e.g., Heinz \& Grimm 2005). We consider the estimated mass of the central SMBH to be $M_{\rm BH} \approx 3.5 \times 10^9 M_{\odot}$ (Walsh et al. 2013). Here, it is to be noted that for hot mode accretion feeding the host SMBHs, the mass inflow rate ($\dot M$) is often inferred to get diminished with decreasing radius (e.g., Fabian et al. 1982; Nulsen et al. 1984; Stewart et al. 1984; Quataert \& Narayan 2000). For a simple logarithmic galactic potential adopted in the study of Quataert \& Narayan (2000), for M87, $\dot M$ has been inferred to decrease from $\sim 1 \,  M_{\odot} \, {\rm yr^{-1}}$ at a few kiloparsecs to $\sim 0.3 \, M_{\odot} \, {\rm yr^{-1}}$ on scales $\lesssim 300$ parsecs where it roughly remains constant. Following Quataert \& Narayan (2000), and for an outflow efficiency of $\sim 10 \, \%$ [e.g., Churazov et al. 2005; Ghosh \& Bhattacharya 2019 (submitted)], one can make a rough estimate of the power associated with these shocks. For $\Delta \mathscr{E}$ in the range of $\approx (10^{-3} - 10^{-4}) \, \% $ (see Fig. \ref{Fig9}), this yields a value of $\approx 1.8 \times (10^{40} - 10^{39}) \, {\rm ergs \, s^{-1}}$ for M87. Interestingly, this range of the value is roughly the same order of magnitude as the core radio power. This dissipated energy could be potentially released as a flare from the flow. Even if one avoids all the uncertainties to estimate $\Delta \mathscr{E}$ in the context of M87 jets, 
$\Delta \mathscr{E}$ is not expected to be less than $\approx (10^{39} - 10^{38}) \, {\rm ergs \, s^{-1}}$. Nonetheless, the plausible scenario we discussed here, is an oversimplification of the real situation, and whether such shocks (even if possible) could actually be associated with flaring in outflows/jets is hard to predict at this point.  

One of the notable contributions of the galactic potential that we found from our study (also see RGJ18) is its tendency to enhance the mass inflow rate. 
While it is being widely held that Bondi-type accretion could provide requisite mass supply rate to power low-luminous or low-excitation galactic nuclei, where in fact, a tight correlation is being observed between estimated Bondi accretion rate and jet kinetic power in many massive ellipticals (e.g., Allen at al. 2006; Nemmen et al. 2007; Fujita et al. 2016), few authors have however argued that this may not be sufficient enough to fuel more powerful and luminous radio sources (e.g., Rafferty et al. 2006; Russell et al. 2013). Narayan \& Fabian (2011) had discussed plausible scenarios, for instance, if one incorporates more realistic effects like convection, magnetohydrodynamic (MHD) effects, or thermal conduction, mass supply rate in the context of spherical/quasi-spherical accretion may get considerably suppressed (also see \S I in this context). In fact, as a consequence of MHD effects, mass accretion rate may get reduced by orders of magnitude, even in the case of a spherical Bondi flow (e.g., Igumenshchev \& Narayan 2002; Igumenshchev 2006). It has been pointed out by Narayan \& Fabian (2011) that if such suppression of mass accretion rate occurs in actual reality, one may find it quite baffling to explain how the observed jet power could track the estimated Bondi rate. This may then raise a relevant question (as also pointed out by other authors) as to whether hot mode accretion could indeed power these massive galaxies, or whether one should invoke alternative fueling mechanisms (e.g., Pizzolato \& Soker 2005; Gaspari et al. 2013; Voit \& Donahue 2015). In this 
context, we would like to point out that if one incorporates the effect of galactic potential in the flow, as our findings indicate, such effect of galactic potential could 
possibly lead to an enhancement of mass supply rate, which could then potentially counteract the plausible suppressive effects. Nonetheless, here one can only make a speculative guess and whether such effect of galactic potential could indeed provide enough fuel enabling hot mode accretion to power these galaxies, requires further investigations; a particularly interesting case would be to conduct numerical/analytical studies of such spherical-type flow by incorporating the MHD effects in the context of our five-component galactic system. Another interesting consequence of this enhancement of mass accretion rate, is that, the radiative cooling might then become important in the spherical Bondi-type flow. It would be then interesting 
to include radiative losses in the flow (see e.g., Mathews \& Guo 2012), or incorporate some simple radiative transfer scheme, for instance, as being used by Chakrabarti \& Sahu (1997), to check how the radiative cooling would impact the dynamics of spherical accretion. Such a study would be pursued somewhere else in the future. 

Finally, we would like to end our discussion with a small caveat: In the present study, to describe our five-component galactic system, we have considered a generalized 
NFW-type profile to describe DM halo, or a generalized type Sersic-profile to describe stellar mass distribution, or the standard $\beta$-profile to describe hot gas mass 
distribution, and our analysis suggests that the effect of galactic potential on different classes of polytropic flows, remains largely consistent. However, a number of other galactic mass-component models are also available in literature, for e.g., for DM models one can see the following references (e.g., Merritt et al. 2006; Graham et al. 2006; Memola et al. 2011; Stuchl\'ik \& Schee 2011); for stellar component, one can have e.g., de Vaucouleurs profile, or Hernquist profile (Hernquist 1990); for hot gas distribution, empirical models such as double $\beta$-model (e.g., Mohr et al. 1999) or modified $\beta$-model (Vikhlinin et al. 2006) also exist. It would be then worthy to examine the behavior of such spherical/quasi-spherical flows in presence of them, to check whether the main findings of our work remain consistent with other galactic models. In the foreseeable future we would wish to undertake such a study. Secondly, our analysis indicates that for a generic class of polytropic flows, galactic potential induced shocks may be quite common in spherical wind-type outflowing solutions. However, realistic outflows/jets are collimated sturctures, and it would be then necessary to explore the possibilities of such mode of shock transitions in more collimated type outflowing solutions.

\appendix
\section{Five-component galaxy model}

Considering elliptical galaxies to have an approximately spherically symmetric mass distribution, in the presence of repulsive cosmological constant $\Lambda$, one can appropriately describe the spacetime geometry exterior to such mass distribution by Schwarzschild-de Sitter (SDS) metric. Adopting the approach in Stuchl\'ik \& Schee (2011), RGJ18 obtained the net galactic force function for our five-component galaxy model using an appropriate pseudo-Newtonian potential (PNP) corresponding to SDS geometry prescribed by Stuchl\'ik et al. (2009), which can simply be represented as a sum of all the individual force terms associated with various gravitational components of the host galaxy, which follows (RGJ18)

\begin{align}
\mathscr{F}_{\rm Gal} \, (r) =  \mathscr{F}_{\rm SMBH} \, (r) + \mathscr{F}_{\rm star} \, (r) + \mathscr{F}_{\rm DM} \, (r) + \mathscr{F}_{\rm gas} \, (r) + \mathscr{F}_{\Lambda} \, (r) \, , 
\label{A1}
\end{align}
where, $\mathscr{F}_{\Lambda} = - {\Lambda c^2 \, r}/3$ is the corresponding force term associated with $\Lambda$ ($\Lambda \sim 10^{- 52} \, {\rm m}^{- 2}$). In the above framework, one would then be able to deal with the relevant astrophysical phenomena inside the galactic halo, and also outside the mass distribution of the galaxy (for details see Stuchl\'ik \& Schee 2011). 

RGJ18 obtained the force functions associated with different mass components of the elliptical galaxy, following Mamon \& Lokas (2005b; ML05b). The force functions are quite complicated and have been elaborately described in RGJ18, we do not repeat it here. For DM distribution, we considered the generalized version of NFW type profile prescribed by 
Jing \& Suto (2000; hereinafter `JS') that provides a double power-law with an arbitrary value of inner slope $- \gamma; 1 \, \lsim \, \gamma \, \lsim \, 3/2$. 
Here, following RGJ18, we adopt both limits of $\gamma$: $\gamma = 1$ (the usual NFW model), and $\gamma = 3/2$ (referred to as JS-3/2 model). To account for the gravitational force function associated with the surrounding hot X-ray emitting gaseous medium, as a first approximation, we assumed the so-called standard $\beta$-profile (Cavaliere \& Fusco-Femiano 1976, 1978; Sarazin 1986) to describe the surrounding gas distribution (RGJ18, ML05b), which can reasonably well represent the overall structure of the X-ray emitting hot gas in clusters of galaxies and massive ellipticals (see e.g., Xu et al. 1998; Akahori \& Masai 2005; Arnaud 2009).

As being discussed in RGJ18, in order to quantify $\mathscr{F}_{\rm Gal} \, (r)$, it is necessary to specify the values of the  
following four quantities: $M_\nu$, $\mathit{f}_{\rm star}$, $\mathit{f}_{\rm DM}$, $\mathit{f}_{\rm gas}$, where $M_\nu$ is the total mass within the virial radius, 
$\mathit{f}_{\rm star}$, $\mathit{f}_{\rm DM}$ and $\mathit{f}_{\rm gas}$ are the mass fractions of stellar, DM and hot gas within the virial 
radius, respectively. Furthermore, to evaluate these quantities, one needs to provide the values of four parameters: $L_B$ (blue-band luminosity of the galaxy), 
$\Upsilon_{\star\, B}$ (stellar mass-to-light ratio in the blue-band), $\Upsilon_B$ (galactic mass-to-light ratio), and 
$\mathit{f}_b$ (baryon mass fraction within the virial radius); note that mass-to-light ratio is always expressed in the units of 
$M_{\odot}/L_{\odot}$, where $M_{\odot}$ is the solar mass and $L_{\odot}$ is the solar luminosity, respectively. Thus if one specify $\Upsilon_{\star\, B}$ and $L_B$, the estimate 
of galactic force function $\mathscr{F}_{\rm Gal} \, (r)$ would then simply depend on two parameters: galactic mass-to-light ratios (${\Upsilon_B}$), and baryon mass fraction 
of the galaxies (${\mathit{f}_b}$) \footnote{For the expressions for $M_\nu$, and stellar, DM and hot gas mass fractions, see section 2 in RGJ18.}. 

Following e.g., ML05b, Capelo et al. (2010), RGJ18, here, we choose a fiducial value of blue-band luminosity $L_B \simeq 2 \times 10^{10} \, L_{\odot}$. Such a choice is motivated 
by the fact that this value of $L_B$ roughly resembles ``the luminosity at the break of the field galaxy luminosity function $L^{\star}_B = 1.88 \times 10^{10} \, L_{\odot}$'' 
(e.g., Mamon \& Lokas 2005a). Again, following the similar authors, here, we choose the fiducial value of $\Upsilon_{\star\, B} = 6.5$. It is hard to estimate precise 
values of ${\Upsilon_B}$ for individual elliptical galaxies from direct measurements. In the analysis of RGJ18, following ML05b, the present authors chose four fiducial values of 
$\Upsilon_B \simeq (14, 33, 100, 390)$, which were observationally motivated. The typical choice of $\Upsilon_B \simeq 390$ corresponds to mass-to-light
ratio of the Universe (${\overline{\Upsilon}}_{B}$), whereas the lower bound of $\Upsilon_B \simeq 14$ actually corresponds to negligible or no DM content scenario. 
For the present purpose, we study our system for this wide range of $\Upsilon_B \simeq (14-390)$, so as to accommodate a wide possible galactic systems. Nonetheless, here we treat 
$\Upsilon_B$ as a free parameter and investigate our system with a wide variation of $\Upsilon_B$ (particularly, for the range of $100 \lesssim \Upsilon_B \lesssim 390$), to examine the possible dependence of shock related quantities on this galactic parameter. For different choices of ${\Upsilon_B}$, one can asses ${\mathit{f}_b}$. 
For $100 \lesssim \Upsilon_B \lesssim 390$, following ML05b and RGJ18, one can consider $\mathit{f}_b = \overline{\mathit{f}}_b \simeq 0.14$, where $\overline{\mathit{f}}_b$ is the `mean baryon mass fraction of the Universe'. For $\Upsilon_B  \simeq  33$, again, following the similar authors, we estimate $\mathit{f}_b \simeq 0.4243$. Note that for $\Upsilon_B \simeq 14$ with no DM case, $\mathit{f}_b \simeq 1$. Using these values of $\mathit{f}_b$, one then can quantify $\mathscr{F}_{\rm Gal}$ for different choices of $\Upsilon_{B}$, required for our numerical computation. 

\end{document}